\documentclass[journal]{IEEEtran}

\usepackage{amsfonts}
\usepackage{amssymb}
\usepackage{stfloats}
\usepackage{cite}
\usepackage{graphicx}
\usepackage{epstopdf}
\usepackage{psfrag}
\usepackage{subfigure}
\usepackage{amsmath}
\usepackage{array}
\usepackage{stfloats}
\usepackage{multirow}
\usepackage{color}
\usepackage{booktabs}
\usepackage{url}
\usepackage{graphicx}
\usepackage{enumitem}
\usepackage{lipsum} % 调用跨栏长公式宏包
\usepackage{algorithm}
\usepackage{algpseudocode}
\usepackage{marvosym} 
\usepackage{url}
\usepackage{hyperref}

\newtheorem{Thm}{Theorem}

\newtheorem{Def}{Definition}

\newtheorem{Prop}{Proposition}

\newcommand{\tabincell}[2]{
    \begin{tabular}{@{}#1@{}}
        #2
    \end{tabular}
}

\begin{document}

\title{A Novel Energy Efficiency Metric for Next Generation Wireless Communication Networks}

\author{Tao~Yu, Shunqing~Zhang,~\IEEEmembership{Senior Member, IEEE,}

Xiaojing~Chen,~\IEEEmembership{Member, IEEE,} and Xin~Wang,~\IEEEmembership{Senior Member, IEEE}
\thanks{This work was supported by the National Natural Science Foundation of China (NSFC) under Grants 62132013, 62071284 and 61901251, the Innovation Program of Shanghai Municipal Science and Technology Commission under Grants 21ZR1422400, 20JC1416400 and 20511106603, and Key-Area Research and Development Program of Guangdong Province under Grant 2020B0101130012.}
\thanks{Tao Yu, Shunqing Zhang, and Xiaojing Chen are with Shanghai Institute for Advanced Communication and Data Science, Key laboratory of Specialty Fiber Optics and Optical Access Networks, Shanghai University, Shanghai, 200444, China (e-mails: \{yu\_tao, shunqing, jodiechen\}@shu.edu.cn).

Xin Wang is with the Key Laboratory for Information Science of Electromagnetic
Waves (MoE), Department of Communication Science and Engineering, Fudan University, Shanghai 200433, China (e-mail: xwang11@
fudan.edu.cn).}
\thanks{Corresponding Author: Shunqing Zhang.}
}

\markboth{Journal of \LaTeX\ Class Files,~Vol.~14, No.~8, August~2015}%
{Shell \MakeLowercase{\textit{et al.}}: Bare Demo of IEEEtran.cls for IEEE Journals}

\maketitle

\begin{abstract}
As a core performance metric for green communications, the conventional energy efficiency definition has successfully resolved many issues in the energy efficient wireless network design. In the past several generations of wireless communication networks, the traditional energy efficiency measure plays an important role to guide many energy saving techniques for slow varying traffic profiles. However, for the next generation wireless networks, the traditional energy efficiency fails to capture the traffic and capacity variations of wireless networks in temporal or spatial domains, which is shown to be quite popular, especially with ultra-scale multiple antennas and space-air-ground integrated network. In this paper, we present a novel energy efficiency metric named {\em integrated relative energy efficiency} (IREE), which is able to jointly measure the traffic profiles and the network capacities from the energy efficiency perspective. On top of that, the IREE based green trade-offs have been investigated and compared with the conventional energy efficient design. Moreover, we apply the IREE based green trade-offs to evaluate several candidate technologies for 6G networks, including reconfigurable intelligent surfaces and space-air-ground integrated network. Through some analytical and numerical results, we show that the proposed IREE metric is able to capture the wireless traffic and capacity mismatch property, which is significantly different from the conventional energy efficiency metric. Since the IREE oriented design or deployment strategy is able to consider the network capacity improvement and the wireless traffic matching simultaneously, it can be regarded as a useful guidance for future energy efficient network design.  
\end{abstract}


\section{Introduction} \label{sect:intro}
\IEEEPARstart{E}{nergy} efficiency (EE) has been extensively studied during the past several decades \cite{zhang2019first, li2011energy}. By dividing the overall system throughput over the total energy consumption, conventional EE concept \cite{xu2012energy} captures two most important performance measures of wireless communication systems, which has been approved as a new metric for the fifth generation (5G) wireless networks \cite{buzzi2016survey}. Based on optimizing the EE related metrics, several energy efficient architectures \cite{ishii2012novel, kaur2015energy}, transmission strategies \cite{dai2016energy}, as well as resource allocation schemes \cite{munir2016energy, li2018energy} have been proposed, and the resultant EE performance has been improved by tens or hundreds times \cite{7939957}.

With the vision to connect everything worldwide via nearly instantaneous, reliable, and unlimited wireless resources \cite{giordani2020toward}, the concept of next generation (6G) wireless communication networks has been raised in many existing literature, and several representative technologies have been widely discussed \cite{rappaport2019wireless, letaief2019roadmap} to fulfill this target. For example, in the spatial domain, {\em space-air-ground integrated network (SAGIN)} \cite{liu2018space-air-ground} has been proposed to incorporate terrestrial, airborne, and satellite networks and provides three dimensional coverage for more efficient and reliable connections, while in the frequency domain, {\em Terahertz (THz)} \cite{8663550} and {\em visible light communication (VLC)} \cite{matheus2019visible} are shown to be promising for Tb/s level throughput with ultra-scale multiple antennas and over tens of gigahertz (GHz) available bandwidth. Massive and heterogeneous transmit entities and available frequency bands have raised several technical challenges in evaluating energy related performance, and the existing EE metric may not be able to reflect the unique characteristics of 6G networks due to the following reasons.

\begin{itemize}
    \item{\em 3D Coverage Extension}. In the past five generations of wireless networks, the traditional EE metric is sufficient to characterize the energy efficient properties of terrestrial networks with two dimensional coverage requirements. For future 6G networks with integrated non-terrestrial networks (NTN), such as satellite or unmanned aerial vehicle (UAV) communications, the three dimensional non-homogeneous coverage features can hardly be described by the existing EE framework. This is partially because the current EE definition ignores the non-uniform service capabilities in the vertical dimension. 
    \item{\em Dynamic Network Capability and Traffic Variations}. With NTN network entities, the instantaneous network capability is no longer static due to the satellite movement, dynamic UAV's trajectory or other mobility issues. Meanwhile, with ultra-scale multiple antennas and ultra-wide frequency bands as illustrated before, the temporal and spatial dynamics of network capability become much more complicated than the existing wireless networks. Hence, a brand-new network EE definition to incorporate network temporal and spatial variations is required, especially when coupled with wireless traffics from ultra-fast mobile terminals. 
\end{itemize}

To address the above issues, we propose a novel EE performance metric, namely {\em Integrated Relative Energy Efficiency (IREE)}, to evaluate 6G wireless networks. Different from the conventional EE definition, we measure the divergences between the time-varying 3D network capability and traffic requirements, and incorporate them into the IREE metric. Through this definition, we are able to describe the network EE under advanced SAGIN architecture and heterogeneous wireless environments. In addition, we also analyze the impacts on the fundamental green trade-offs as proposed in \cite{chen2011fundamental} and the potential benefits on designing energy efficient 6G network architectures and transmission techniques.

\subsection{Related Works} \label{subsect:related}

The definition of EE has been extensively studied over the past several decades \cite{chen2010energy, zhao2013efficiency, zhao2011radio}. For example, the bit-per-Joule EE metric was first proposed in \cite{goldsmith2002design}, which evaluates the entire transmission capacity over the corresponding energy consumption. Area energy efficiency (AEE) was later defined in \cite{wang2010energy} to measure EE under different network topology with heterogeneous coverage, where bit-per-Joule per unit area is the common metric. In order to calibrate the absolute performance of different wireless networks, energy consumption rating (ECR) \cite{5783984} has been proposed as the ratio of energy consumption over effective system capacities. Instead of using network throughput, another type of EE metrics apply spectrum efficiency (SE) as the system utility, and the related EE is defined to be SE over the entire power consumption. Typical examples include power efficiency (b/s/Hz/W) and radio efficiency (b$\cdot$m/s/Hz/W) \cite{zhao2011radio}. 

With the above EE definition, energy efficient transmission schemes for next generation wireless networks have been widely studied quite recently. In SAGIN architecture, an energy efficient transmission scheme to adaptively select terrestrial relaying has been proposed in \cite{ruan2018energy} under symbol error rate constraints, and extended to maximize the effective capacity bound for EE enhancement in \cite{ruan2019energy}. In \cite{zeng2017energy-efficient} and \cite{huang2020energy}, UAVs have been selected as the communication entities, while energy efficient trajectory optimization and beamforming schemes are the main approaches to maximize the famous {\em bit-per-Joule} EE metric. Another common approach for EE improvement is to directly reduce energy or power consumption with quality-of-service (QoS) or other performance constraints \cite{sodhro2020towards, zhen2020energy}. As key concerns for energy efficient 6G Internet-of-Things (IoT) networks, access delay \cite{verma2020towards} and security constraints \cite{kaur2021secure} have been guaranteed when the overall power consumption is minimized.

Due to the dual mobility of SAGIN enabled 6G networks and highly dynamic temporal-spatial wireless transmission environments as illustrated before, traditional EE usually suffers from evaluating time-varying and 3D utility functions. To the best of our knowledge, the temporal-spatial wireless traffic and network capacity mismatch issue has not been considered in EE definition and evaluation, and a new EE definition for 6G networks is thus required.

\begin{figure}[t] %[h]
\centering  
\includegraphics[height=8cm,width=8cm]{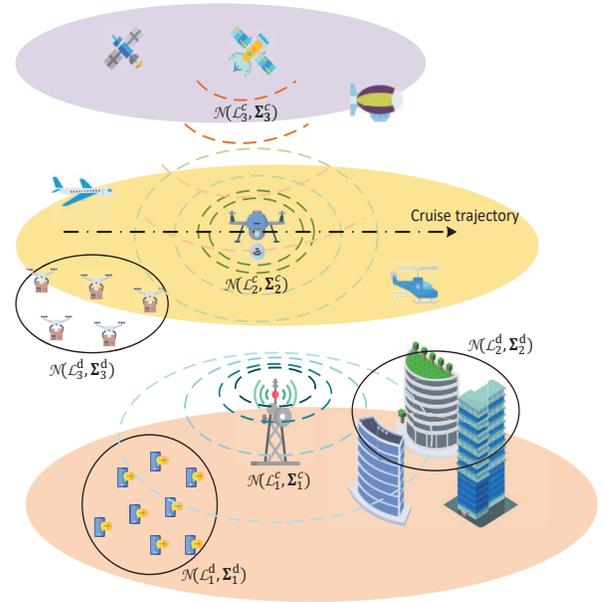}
\caption{An illustrative example of 3D wireless traffic and network capacity distributions for 6G networks. The wireless traffic is modeled as a 3-order Gaussian mixture model, where the corresponding parameters are given as, $\omega^d_1 = \omega^d_2 = \omega^d_3 = 1/3$, $\mathcal{L}^{d}_1 = (200,200,10), \mathcal{L}^{d}_2 = (800,200,400), \mathcal{L}^{d}_3 = (600,800,800)$, and $\mathbf{\Sigma}^d_1 = \text{diag} (40000,40000,25600), \mathbf{\Sigma}^d_2 = \text{diag} (40000,10000,25600), \mathbf{\Sigma}^d_3 = \text{diag} (10000,40000,22500)$. }
\label{fig:scenario}
\end{figure}

\subsection{Contribution \& Organization} \label{subsect:contribution}
In this paper, IREE is proposed as a new EE metric for 6G network and the main contributions of this paper are summarized as follows.

\begin{itemize}
    \item {\em IREE Metric:} In order to address 3D coverage extension and dynamic network capability/traffic variations, we propose a novel EE metric by jointly considering the heterogeneous path loss effects of wireless propagation environment, the temporal and spatial network capability imbalance, as well as the mismatched network capacity and traffic distributions. Inspired by the metrics of probability distribution similarity, the proposed IREE metric is able to evaluate the non-uniform network capacity/traffic variations, and the effectiveness of 3D SAGIN coverage in EE aspect.
    \item {\em IREE based Green Trade-offs:} With the proposed IREE, we extend the green fundamental trade-offs framework as defined in \cite{chen2011fundamental}, and compare with the conventional EE metric in order to show the effectiveness under 6G SAGIN architecture. Specifically, we show that the optimal energy efficient scheme should be adaptive with different traffic distributions for IREE based green trade-offs, while it remains static for conventional SE-EE or deployment efficiency (DE)-EE trade-offs.
    \item {\em Impacts on Energy Efficient 6G Network Design:} By utilizing the IREE based green trade-offs, we extensively evaluate the performance gain in terms of IREE for several 6G technologies, including Reconfigurable Intelligent Surfaces (RIS) and SAGIN architecture. Based on that, we are able to figure out more energy efficient design strategies to achieve better IREE performance, which are significantly different from the conventional EE oriented schemes, such as \cite{miao2009cross} and \cite{5503900}.
\end{itemize}

The remaining part of the paper is organized as follows. The background information about 6G network architecture, power models, as well as the conventional EE definitions are introduced in Section~\ref{sect:background}. In Section~\ref{sect:new_definition}, we propose the novel IREE definition and show the benefits over conventional EEs under different traffic variations. The extension to fundamental green trade-off framework is then discussed in Section~\ref{sect:trade-off}, followed by some analytical results on energy efficient 6G network design in  Section~\ref{sect:impacts_EE}. In Section~\ref{sect:num_res}, we provide some numerical examples on the IREE based green trade-offs. Finally, concluding remarks are provided in Section~\ref{sect:conc}.

\section{Backgrounds} \label{sect:background} 
In this section, we briefly introduce the utility function and the power model of terrestrial and non-terrestrial networks, followed by some conventional definitions of EE metrics.

\subsection{Utility and Power Models}

As mentioned before, the conventional EE is often defined to be the overall utility values over the entire power or energy consumption, where the utility function could be capacity, throughput, or SE. For illustration purpose, we choose Shannon capacity \cite{shannon1948mathematical} as the utility function, and the mathematical expression for SAGIN is given by,
\begin{equation} 
\label{eq:capacity}
C^{T/A/S} = B^{T/A/S} \log_2 \left (1 + \frac{ P_t/L^{T/A/S}}{I^{T/A/S} + B^{T/A/S} N_0} \right)
\end{equation}
where $B^{T/A/S}$, $L^{T/A/S}$, and $I^{T/A/S}$ denote the bandwidth, the normalized path loss coefficients, and the generated interference of terrestrial, airborne and satellite base stations, respectively. $P_t$ and $N_0$ represent the normalized transmit power and noise spectral density, respectively.

The power consumption of SAGIN networks is composed of the following three parts, i.e., the hovering power, the static power, and the dynamic power. Denote $P^{A/S}_{m}$, $P^{T/A/S}_{i}$, $P^{T/A/S}_c$ to be the hovering power\footnote{We omit the superscript $T$, since terrestrial networks can not move in general.}, the idle mode power consumption, and the circuit power consumption, respectively. The entire power consumption, $P^{T/A/S}$, is thus given by,
\begin{eqnarray} 
\label{eq:power_consumption_tas}
P^{T/A/S} &=& P^{A/S}_{m} + Pr_{i} P^{T/A/S}_{i} + \left(1 - Pr_{i}\right) \times \Big[ \lambda^{T/A/S} \nonumber\\ 
&&  P^{G/A/S}_{t} + P^{T/A/S}_c \Big],
\end{eqnarray}
where $Pr_{i}$ represents the probability of idle mode and $\lambda^{T/A/S}$ is the efficiency of RF amplifiers.

\subsection{Conventional Definition of EE}

By summarizing the utility and power consumption of SAGIN, we can obtain the instantaneous utility $C_T(\tau)$ and power consumption $P_T(\tau)$ in the time stamp $\tau$ as follows.
\begin{eqnarray}
C_T(\tau) & = & C^T + C^A + C^S, \label{eqn:tot_cap}\\
P_T(\tau) & = & P^T + P^A + P^S.
\end{eqnarray}
Following the conventional definitions of EE and AEE as specified in \cite{shahab2015assessment}, we can define EE and AEE of SAGIN similarly as follows.
\begin{Def}[EE and AEE of SAGIN]
\label{def:ee}
The EE and AEE of SAGIN networks are given by,
\begin{eqnarray}
\eta_{EE} & = & \frac{C_T(\tau)}{P_T(\tau)}, \\
\eta_{AEE} & = &  \frac{\eta_{EE}^{\max}}{V_\mathcal{A}},
\end{eqnarray}
where $V_\mathcal{A}$ denotes the coverage area defined by region $\mathcal{A}$, and $\eta_{EE}^{\max}$ denotes the achievable EE utility within the corresponding coverage.
\end{Def}

The conventional EE and AEE definition for SAGIN is able to evaluate terrestrial networks with heterogeneous architectures and low mobility. However, for non-terrestrial networks, the above EE and AEE may not be suitable due to the following three reasons. First, the Shannon capacity based network utility of high mobility non-terrestrial base stations (BSs) is unable to describe the dynamic variations of network capabilities. Second, the wireless traffic variations have not been considered in the above EE and AEE definition, which is not suitable for many 6G applications with heterogeneous traffic distributions. Last but not the least, the extension to 3D SAGIN coverage is {\em not} straightforward for the conventional EE and AEE, where a novel metric for measuring the 3D network utility is called for.

\section{Proposed IREE Metric} \label{sect:new_definition} 

In this section, we introduce the IREE metric, which is more convenient for dynamic network capability/traffic variations and 3D coverage extension. In what follows, we compare the proposed IREE with other conventional EE metrics.

\subsection{IREE Definition}
Without loss of generality, we denote variables $D_{T}(\mathcal{L}, \tau)$ and $C_{T}(\mathcal{L}, \tau)$ to be the time-varying data traffic and network capacity at the location $\mathcal{L}$ and the time stamp $\tau$, respectively. When the data traffic is greater than the network capacity, i.e., $D_{T}(\mathcal{L}, \tau) \geq C_{T}(\mathcal{L}, \tau)$, the achievable throughput is limited by the network capacity $C_{T}(\mathcal{L}, \tau)$, and vice versa. Therefore, a more reasonable utility measure for given $\mathcal{L}$ and $\tau$ can be expressed as, $\min\{C_{T}(\mathcal{L}, \tau), D_{T}(\mathcal{L}, \tau)\}$. Following the conventional EE definition, we have the {\em integrated EE} form for time-varying 3D coverage as follows,
\begin{eqnarray}\label{def:iee}
\eta_{IEE} = \frac{\int_{\tau}\iiint_{\mathcal{A}}\min\{C_{T}(\mathcal{L}, \tau), D_{T}(\mathcal{L}, \tau)\} \textrm{d}\mathcal{L} \textrm{d}\tau}{\int_{\tau}P_{T}(\tau)\textrm{d}\tau},
\end{eqnarray}
where we follow equation \eqref{eqn:tot_cap} to have the time-varying network capacity as $C_{T}(\mathcal{L}, \tau) = C^{T}(\mathcal{L}, \tau)+C^{A}(\mathcal{L}, \tau)+C^{S}(\mathcal{L}, \tau)$.

In the practical evaluation, to measure the network capacity and the data traffic for each location $\mathcal{L}$ and time stamp $\tau$ is complexity prohibited. To address this issue, we denote the corresponding 3D data traffic and capacity distributions by $f^{d}(\mathcal{L}, \tau)$ and $f^{c}(\mathcal{L}, \tau)$, where the mathematical expressions are given by,
\begin{eqnarray}
f^{d}(\mathcal{L}, \tau) & = & \frac{D_{T}(\mathcal{L}, \tau)}{ \int_{\tau} \iiint_{\mathcal{A}} D_{T}(\mathcal{L}, \tau) \textrm{d}{\mathcal{L}} \textrm{d}\tau }, \label{eq:traffic_distri} \\
f^{c}(\mathcal{L}, \tau) & = & \frac{C_{T}(\mathcal{L}, \tau)}{ \int_{\tau} \iiint_{\mathcal{A}} C_{T}(\mathcal{L}, \tau) \textrm{d}{\mathcal{L}} \textrm{d}\tau}. \label{eq:cap_distri}
\end{eqnarray}
With some mathematical manipulations as shown in Appendix~\ref{appendix:def}, we propose the IREE definition as follows.

\begin{Def}[IREE] \label{def:IREE}
The IREE of wireless networks, $\eta_{IREE}$, is defined to be,
\begin{eqnarray} \label{eq:def_iree}
\eta_{IREE} = \frac{ \min \{C_{Tot}, D_{Tot}\}\Big[ 1  - \xi(f^{c}, f^{d})  \Big] } {\int_{\tau}P_{T}(\tau)\textrm{d}\tau }.
\end{eqnarray}
In the above expression, $C_{Tot} = \int_{\tau}\iiint_{\mathcal{A}}C_{T}(\mathcal{L}, \tau) \textrm{d}\mathcal{L} \textrm{d}\tau$, $D_{Tot} = \int_{\tau}\iiint_{\mathcal{A}}D_{T}(\mathcal{L}, \tau) \textrm{d}\mathcal{L} \textrm{d}\tau$, and $P_{Tot} = \int_{\tau}P_{T}(\tau)\textrm{d}\tau $ denote the total amount of wireless capacity, the total amount of wireless traffic, as well as the total power consumption. $\xi(f^{c}, f^{d})$ is the Jensen-Shannon (JS) divergence, which is equal to $\frac{1}{2} \int_{\tau}\iiint_{\mathcal{A}}  f^{c}(\mathcal{L}, \tau) \log_2 \left[ \frac{2 f^{c}(\mathcal{L}, \tau) }{f^{c}(\mathcal{L}, \tau) + f^{d}(\mathcal{L}, \tau) } \right] \nonumber + f^{d}(\mathcal{L}, \tau)$ $\log_2 \left[ \frac{2 f^{d}(\mathcal{L}, \tau)}{f^{c}(\mathcal{L}, \tau) + f^{d}(\mathcal{L}, \tau)} \right] \textrm{d}\mathcal{L} \textrm{d}\tau$ as defined in \cite{manning1999foundations}.
\end{Def}

From Definition~\ref{def:IREE}, we can realize that there are two approaches to improve the values of IREE. One is to increase the network capacity $C_{Tot}$ as the conventional EE suggested when total traffic need $D_{Tot}$ is large enough, and the other is to decrease the JS divergence $\xi(f^{c}, f^{d})$, i.e., to match the network capacity with the data traffic profiles.

\subsection{Comparison between EE Metrics}
In order to illustrate the differences between IREE and other conventional EE metrics, two types of BSs are generated, including the moving UAV BS and the terrestrial BS located in the center of the evaluation area. For illustration purpose, the transmit power of BSs is 35 dbm and the path loss is generated according to \cite{6334508}. The UAV BS is assumed to cruise along the horizontal direction, and the terrestrial BS remains static. With the above settings, the traffic and network capacity distributions can be approximated by Gaussian models as shown in Fig.~\ref{fig:scenario}.

\begin{figure*}[t]
\centering
\subfigure[]{
\begin{minipage}[c]{0.3\linewidth}
\centering
\includegraphics[height=5.5cm,width=5.5cm]{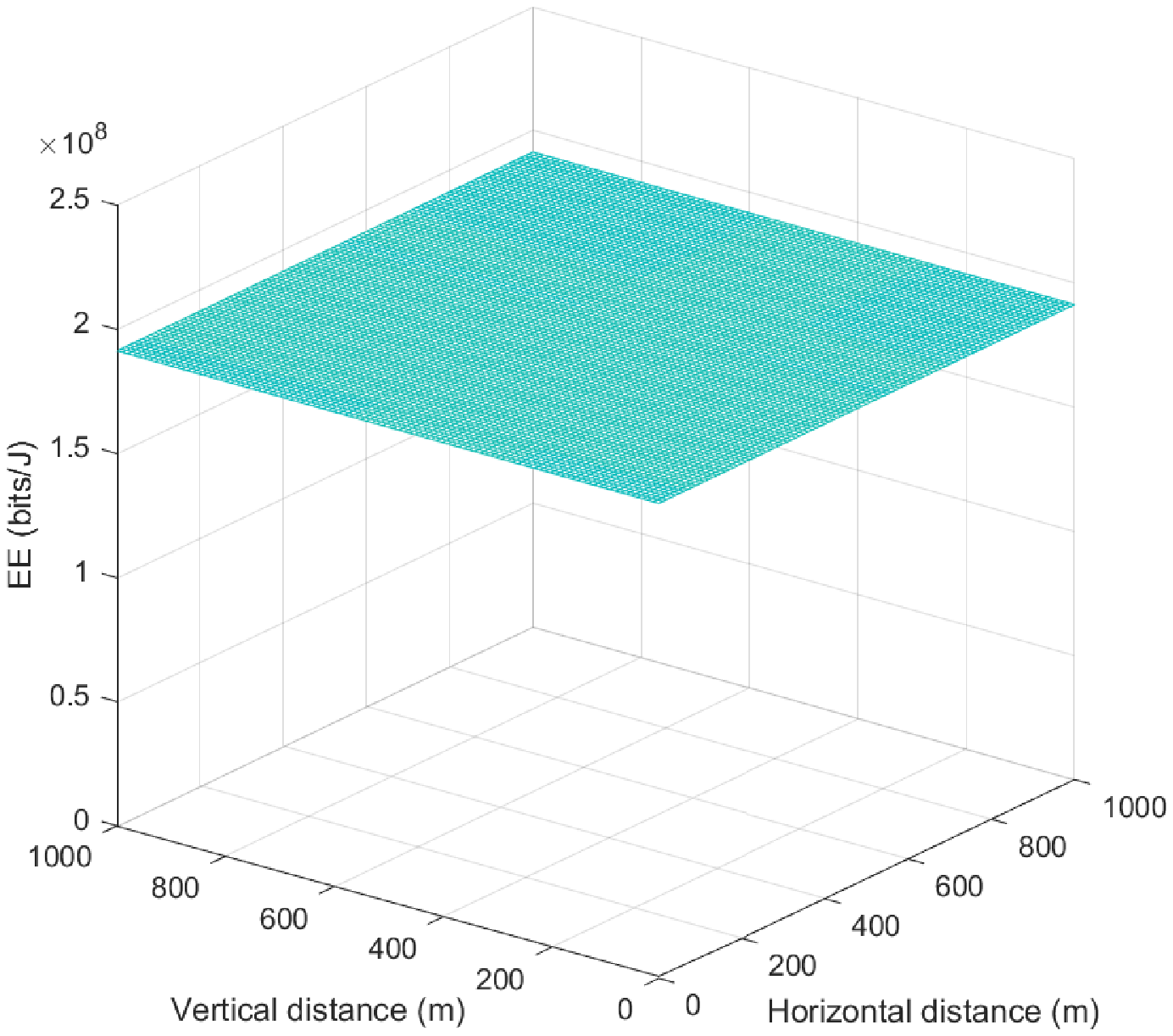}
\label{fig:ee_compare}
% \vspace*{3pt}
\end{minipage}}
\subfigure[]{
\begin{minipage}[c]{0.3\linewidth}
\centering
\includegraphics[height=5.5cm,width=5.5cm]{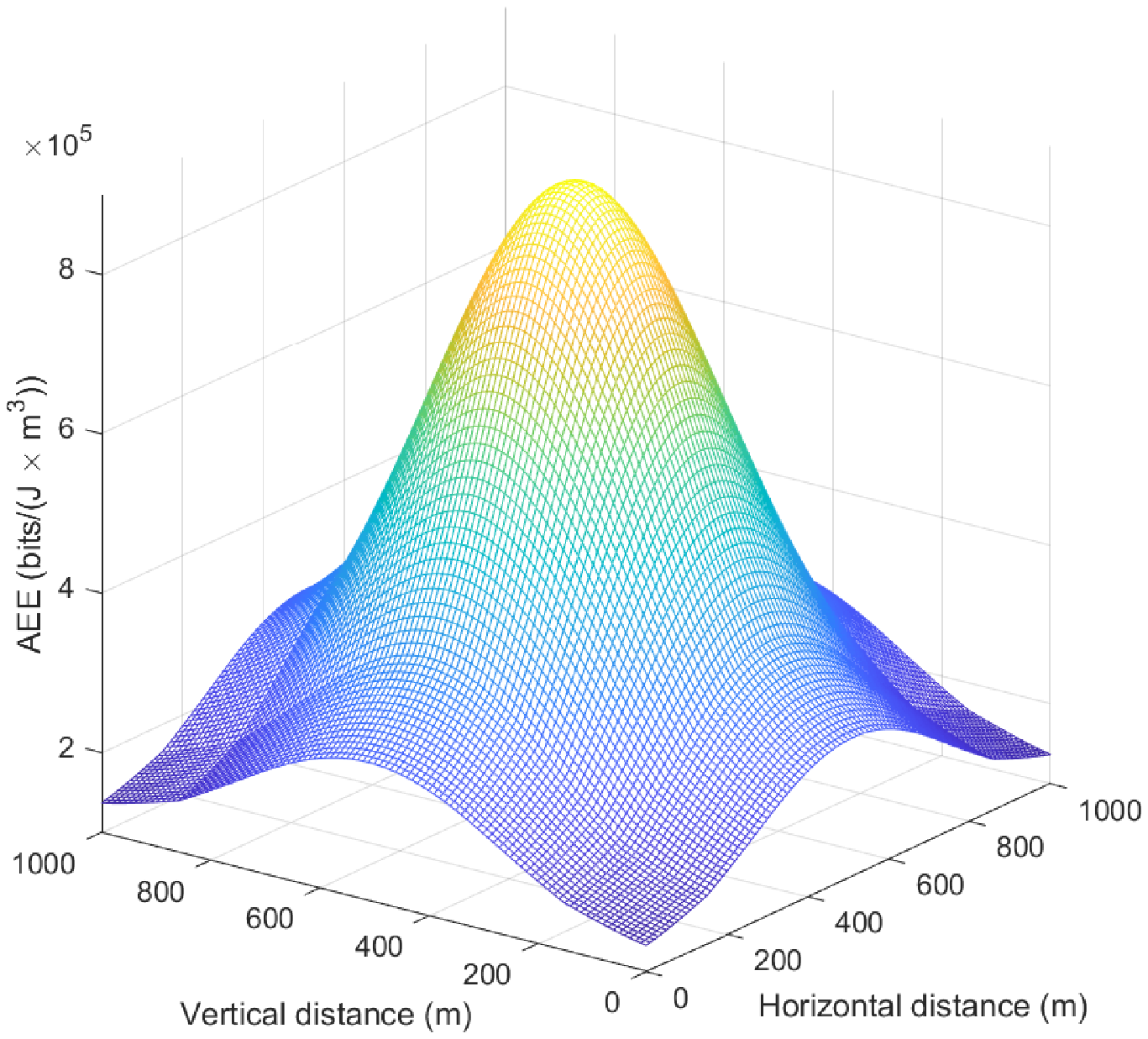}
\label{fig:aee_compare}
% \vspace*{3pt}
\end{minipage}}
\subfigure[]{
\begin{minipage}[c]{0.3\linewidth}
\centering
\includegraphics[height=5.5cm,width=5.5cm]{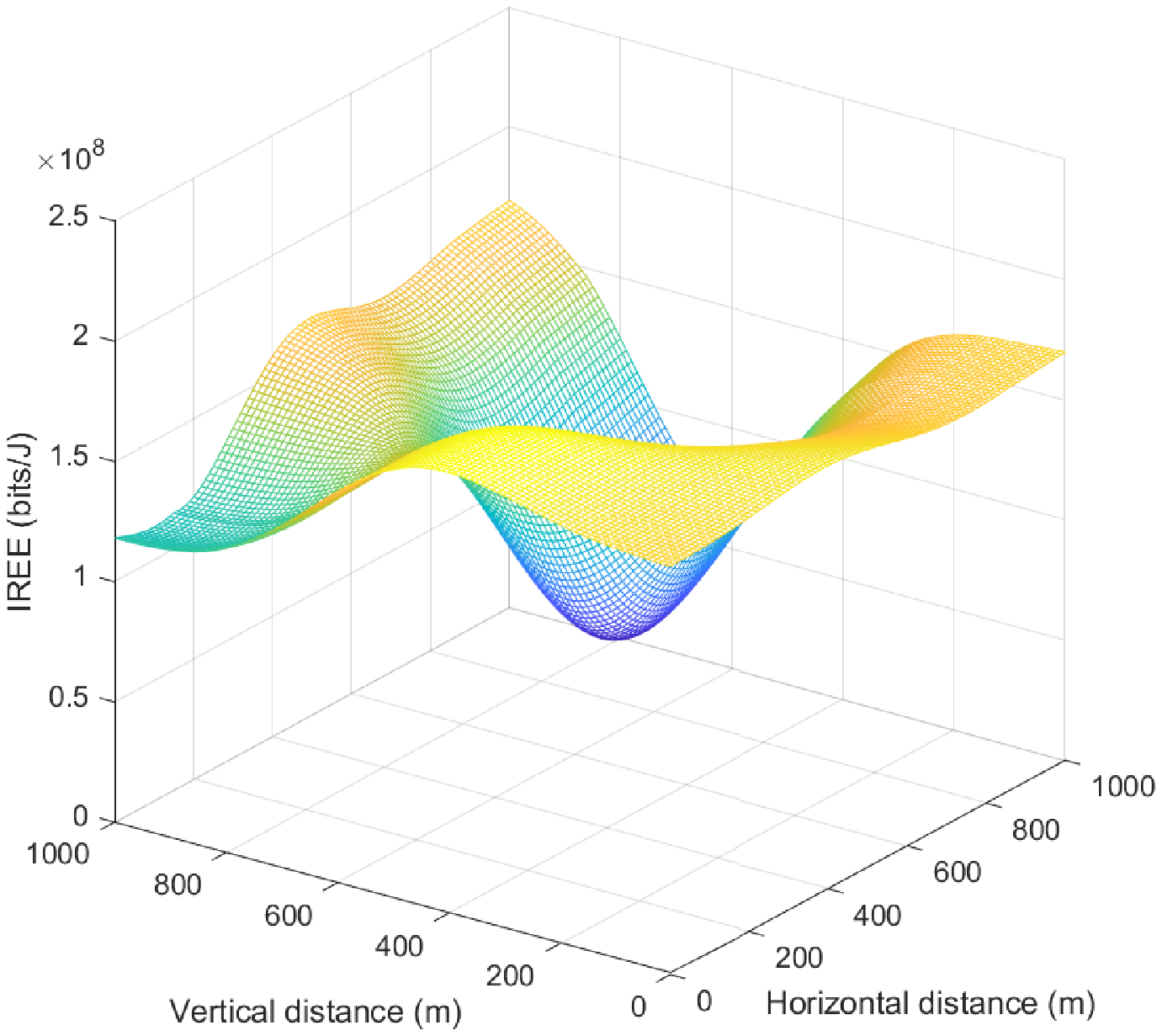}
\label{fig:iree_compare}
% \vspace*{3pt}
\end{minipage}}
% \vspace*{1pt}
\caption{Numerical comparison among EE, AEE and IREE. (a) EE is a smooth surface, which is irrelevant to the traffic distribution. (b) AEE is a ``bell'' shape surface, which scales with the network coverage $V_\mathcal{A}$. (c) IREE is varying with traffic profiles, which is due to the traffic and network capacity mismatch in the horizontal and vertical directions.}
\label{fig:ee_metrics_compare}
\end{figure*}

In Fig.~\ref{fig:ee_metrics_compare}, we compare the proposed IREE metric with the conventional EE and AEE. As shown in Fig.~\ref{fig:ee_compare}, the conventional EE focuses on evaluating the total network capacity over the total power consumption, which is irrelevant to the spatial distribution of the network capability. For AEE, since we take into the consideration of network coverage effects, we can observe an AEE degradation when the network coverage extends. However, both EE and AEE fail to investigate the effects from the non-uniform traffic profiles\footnote{Due to the page limit, the EE, AEE, and IREE comparison under the time-varying traffic profiles is not shown here, which could be obtained straight forwardly. In the rest of this paper, we omit the time stamp $\tau$ to represent the ergodic realization in the time domain.}. Regarding the IREE metric, the traffic profiles have been considered via the JS divergence $\xi(f^{c}, f^{d})$. Therefore, if there is sufficient traffic requirement at the network edge, we can still obtain high IREE value by fully utilizing the network capacity.

\section{Application to Fundamental Green Trade-offs} \label{sect:trade-off}

In this section, we study the IREE based fundamental green trade-offs. Specifically, we derive the IREE versus SE and DE relations, in order to obtain the energy efficient design insights for SAGIN networks. Since the JS divergence $\xi(f^{c}, f^{d})$ is highly impacted by the traffic distribution $f^d(\mathcal{L})$, we assume that it obeys the $K$-order Gaussian Mixture Model (GMM)\footnote{As explained in \cite{8454668, 9220821}, GMM can be applied to fit wireless traffic distribution with appropriate parameters. For illustration purpose, we denote $\mathcal{N}(\mu, \sigma)$ to be the Gaussian random variable with mean $\mu$ and variance $\sigma$, respectively, and $\omega^{d}_{k}$ to be the corresponding weights, such that $\sum_{k=1}^{K} \omega^{d}_{k} = 1$. } as specified in \cite{8454668, 9220821}, i.e., $f^d(\mathcal{L}) = \sum_{k=1}^{K} \omega^{d}_{k} \mathcal{N}(\mathcal{L}^{d}_{k}, \mathbf{\Sigma}^{d}_{k})$, and derive the related green trade-offs in what follows. 

\subsection{SE-IREE Trade-off}
Following the conventional definition of SE, we can derive the average SE of SAGIN networks as follows,
\begin{eqnarray}
\label{def:se}
\eta_{SE} = \frac{ C_{Tot}}{B_{Tot}} = \frac{\int_{\tau}\iiint_{\mathcal{A}}C_{T}(\mathcal{L}, \tau) \textrm{d}\mathcal{L} \textrm{d}\tau}{B^T + B^A + B^S},
\end{eqnarray} 
where $B_{Tot} = B^T + B^A + B^S$ denotes the overall consumed bandwidth of SAGIN networks. Substitute \eqref{def:se} into \eqref{eq:def_iree}, we have the following theorem for the SE-IREE trade-off.

\begin{Thm}[SE-IREE] \label{thm:se-iree} For any given power profile $P_{T}(\tau)$, the SE-IREE relation can be described as
\begin{eqnarray}\label{eq:se_iree_thm}
\eta_{IREE} = \frac{ \min \{B_{Tot} \eta_{SE}, D_{Tot}\}\Big[ 1  - \xi(f^{c}, f^{d})  \Big] }{P_{Tot}}.
\end{eqnarray}
\end{Thm}

As illustrated in \cite{chen2011fundamental}, due to the static power consumption in $P_{Tot}$, the conventional SE-EE trade-off becomes a ``bell'' shape when $P_{Tot}$ varies. This conclusion is based on the assumption that all network capacities can be fully utilized, i.e., $D_{Tot} > C_{Tot}$ and $\xi(f^{c}, f^{d}) = 0 $, and it still holds true when $\xi(f^{c}, f^{d})$ is equal to a constant other than 1. In addition to this effect, Theorem~\ref{thm:se-iree} also reflects that the wireless traffic distribution affects this trade-off relation via the JS divergence, $\xi(f^{c}, f^{d})$. Therefore, we can have two alternative strategies to decrease the value of $\xi(f^{c}, f^{d})$ and improve the network IREE. One is to fine-tune the network capacity distribution via UAV or satellite base stations, and the other is to adjust the wireless traffic distribution via traffic migrations, such as \cite{anwar20215g} and \cite{8967040}. 

Once the network capacity distribution can be modeled as an $L$-order GMM, i.e., $f^c(\mathcal{L}) = \sum_{l=1}^{L} \omega^{c}_{l} \mathcal{N}(\mathcal{L}^{c}_{l}, \mathbf{\Sigma}^{c}_{l})$, we can have the following proposition for the closed-form SE-IREE trade-off.

\begin{figure}[t]
\centering  
\includegraphics[height=7cm,width=7.5cm]{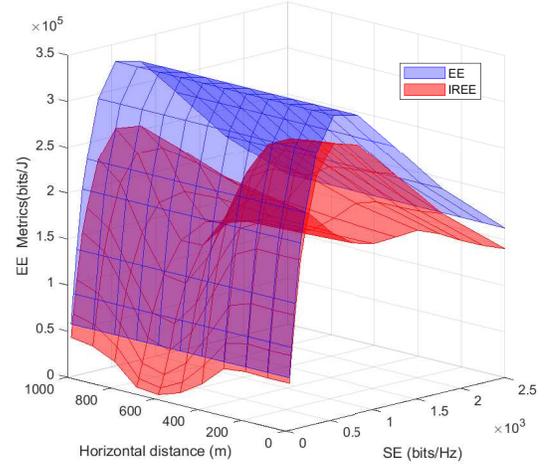}
\caption{The SE-IREE/SE-EE trade-off relations in the horizontal direction. Different from the conventional SE-EE trade-off surface along the horizontal direction, the SE-IREE trade-off surface will be bent down, when the wireless traffic and network capacity mismatch happens.}
\label{fig:iree_punch_holes}
\end{figure}

\begin{Prop}[Closed-Form SE-IREE] \label{prop:se_iree}
For any given GMM based wireless traffic and network capacity distributions, i.e., $f^d(\mathcal{L}) = \sum_{k=1}^{K} \omega^{d}_{k} \mathcal{N}(\mathcal{L}^{d}_{k}, \mathbf{\Sigma}^{d}_{k})$ and $f^c(\mathcal{L}) = \sum_{l=1}^{L} \omega^{c}_{l} \mathcal{N}(\mathcal{L}^{c}_{l}, \mathbf{\Sigma}^{c}_{l})$, the closed-form SE-IREE trade-off relation is given by,
\begin{eqnarray} \label{eqn:SE-IREE}
\eta_{IREE} =\frac{ \min \{B_{Tot} \eta_{SE}, D_{Tot}\} \left[1 - \xi\left(\{\mathcal{L}^{c/d}_{l/k}\},\{\mathbf{\Sigma}^{c/d}_{l/k}\}\right) \right] }{P_{Tot}},
\end{eqnarray}
where the corresponding JS divergence is given by
\begin{equation} \label{eqn:JS_closeform}
\begin{split}
&\xi\big(\{\mathcal{L}^{c/d}_{l/k}\},\{\mathbf{\Sigma}^{c/d}_{l/k}\}\big) = 1 - \frac{1}{2} \times \bigg[\sum_{l = 1}^{L} \omega^c_{l} \times \log_2 \Big( 1 + \\
& \frac{  \sum_{k = 1}^{K} \omega^d_{k} I\big( \mathcal{N}(\mathcal{L}^{c}_{l}, \mathbf{\Sigma}^{c}_{l}) | \mathcal{N}(\mathcal{L}^{d}_{k}, \mathbf{\Sigma}^{d}_{k})\big)}{ \sum_{l' = 1}^{L} \omega^c_{l'} I\big( \mathcal{N}(\mathcal{L}^{c}_{l}, \mathbf{\Sigma}^{c}_{l}) | \mathcal{N}(\mathcal{L}^{c}_{l'}, \mathbf{\Sigma}^{c}_{l'})\big)} \Big) + \sum_{k = 1}^{K} \omega^d_{k} \times \\
& \log_2 \Big ( 1 + \frac{\sum_{l = 1}^{L} \omega^c_{l} I\big( \mathcal{N}(\mathcal{L}^d_{k}, \mathbf{\Sigma}^d_{k}) | \mathcal{N}(\mathcal{L}^c_{l}, \mathbf{\Sigma}^c_{l})\big) }{ \sum_{k' = 1}^{K} \omega^d_{k'} I\big( \mathcal{N}(\mathcal{L}^d_{k}, \mathbf{\Sigma}^d_{k}) | \mathcal{N}(\mathcal{L}^d_{k'}, \mathbf{\Sigma}^d_{k'})\big) } \Big) \bigg].
\end{split}
\end{equation} 
In the above equation, the exponential mutual divergence, $I(\mathcal{N}(\mathcal{L}^{g}, \mathbf{\Sigma}^{g})|\mathcal{N}(\mathcal{L}^{h}, \mathbf{\Sigma}^{h}))$, is equal to $\text{exp} \{-\frac{1}{2} \big[ \ln\frac{|\mathbf{\Sigma}^h|}{|\mathbf{\Sigma}^g|} + \text{Tr}[ (\mathbf{\Sigma}^h)^{-1} \mathbf{\Sigma}^g ] + (\mathcal{L}^h - \mathcal{L}^g)^T (\mathbf{\Sigma}^h)^{-1} (\mathcal{L}^h - \mathcal{L}^g)-3\big]\}$.
\end{Prop}
\IEEEproof
Please refer to Appendix~\ref{appendix:se_iree} for the proof.
\endIEEEproof
From Proposition~\ref{prop:se_iree}, we note that when the total amount of wireless traffic is sufficiently small, IREE will tend to $0$ no matter how much SE is. However, when $D_{Tot}$ is sufficiently large, the SE-IREE trade-off will be highly impacted by the JS divergence $\xi\big(\{\mathcal{L}^{c/d}_{l/k}\},\{\mathbf{\Sigma}^{c/d}_{l/k}\}\big)$. In Fig.~\ref{fig:iree_punch_holes}, we plot the SE-IREE/SE-EE trade-off relations in the horizontal direction, where the mismatched distributions of wireless traffic and network capacity are considered. Different from the conventional SE-EE trade-off surface along the horizontal direction, the SE-IREE trade-off surface will be bent down when this mismatch happens as shown in Fig.~\ref{fig:iree_punch_holes}. It is also worth to mention that Proposition~\ref{prop:se_iree} does not rely on any specific wireless transmission or power consumption models, implying that it can be applied to various wireless scenarios in 6G networks.

\subsection{DE-IREE Trade-off}

Similarly, we can follow the conventional definition of DE and derive the average DE of SAGIN networks as,
\begin{eqnarray}
\label{def:de}
\eta_{DE} = \frac{ C_{Tot}}{W_{Tot}} = \frac{C_{Tot}}{W_{Cap} + \gamma P_{Tot}},
\end{eqnarray} 
where $W_{Tot}$ and $W_{Cap}$ denote the overall deployment cost and the Capital Expenditure (CapEx) of SAGIN networks, respectively. $\gamma$ denotes the Operational Expenditure (OpEx) factor\footnote{For illustration purpose, we assume the OpEx is proportional to the total power consumption $P_{Tot}$, and the following derived results are applicable to other OpEx models as well.}. Substitute \eqref{def:de} into \eqref{eq:def_iree}, we have the following theorem for the DE-IREE trade-off.

\begin{Thm}[DE-IREE] \label{thm:de-iree} For any given power profile $P_{T}(\tau)$, the DE-IREE relation can be described as,
\begin{eqnarray}
\eta_{IREE} &=& \frac{ \min \{ (W_{Cap} + \gamma P_{Tot}) \eta_{DE}, D_{Tot}\} }{P_{Tot}} \nonumber \\
&&\times \Big[ 1  - \xi(f^{c}, f^{d})  \Big].
\end{eqnarray}
\end{Thm}

From Theorem~\ref{thm:de-iree}, we can realize that the network deployment strategy shall consider the conventional DE aspect as well as the JS divergence, $\xi(f^{c}, f^{d})$. In practical wireless communication systems, due to the time varying nature of traffic profiles, UAV BSs are shown to be energy efficient, even with significant hovering energy consumption \cite{mozaffari2017mobile}. This is partially because we can achieve better $\xi(f^{c}, f^{d})$ by adapting the network capacity via UAV BSs, and eventually improve the IREE value.

\begin{Prop}[Closed-Form DE-IREE] \label{prop:de_iree}
For any given GMM based wireless traffic and network capacity distributions, i.e., $f^d(\mathcal{L}) = \sum_{k=1}^{K} \omega^{d}_{k} \mathcal{N}(\mathcal{L}^{d}_{k}, \mathbf{\Sigma}^{d}_{k})$ and $f^c(\mathcal{L}) = \sum_{l=1}^{L} \omega^{c}_{l} \mathcal{N}(\mathcal{L}^{c}_{l}, \mathbf{\Sigma}^{c}_{l})$, the closed-form DE-IREE trade-off relation is given by,
\begin{eqnarray}
\label{eqn:DE-IREE}
\eta_{IREE} &=& \frac{ \min \{ (W_{Cap} + \gamma P_{Tot}) \eta_{DE}, D_{Tot}\} }{P_{Tot}} \nonumber \\
&&\times \Bigg [ 1 - \xi\left(\{\mathcal{L}^{c/d}_{l/k}\},\{\mathbf{\Sigma}^{c/d}_{l/k}\}\right) \Bigg ].
\end{eqnarray}
\end{Prop}
\IEEEproof
Substitute the expression of $\xi\left(\{\mathcal{L}^{c/d}_{l/k}\},\{\mathbf{\Sigma}^{c/d}_{l/k}\}\right)$ into Theorem~\ref{thm:de-iree}, and we have Proposition~\ref{prop:de_iree}. 
\endIEEEproof

From Proposition~\ref{prop:de_iree}, we can conclude that the IREE performance, $\eta_{IREE}$, grows linearly with respect to $\eta_{DE}$ when $D_{Tot}$ is large enough. Meanwhile, the DE-IREE trade-off will also be impacted by the JS divergence $\xi\big(\{\mathcal{L}^{c/d}_{l/k}\},\{\mathbf{\Sigma}^{c/d}_{l/k}\}\big)$, which is more or less the same with the SE-IREE trade-off.

\section{Impacts on Energy Efficient 6G Network Design} 
\label{sect:impacts_EE}

In this section, we apply the obtained IREE based green trade-offs to shed some light on energy efficient 6G network design, especially on the RIS configuration and the SAGIN architecture. 

\subsection{RIS Configuration}

As a candidate technology for 6G networks, RIS configuration \cite{huang2019reconfigurable} can be customized to create virtual line-of-sight (LOS) propagation paths among transceiver pairs with marginal power consumption. Denote $\Delta C^{RIS}_{Tot}$ to be the capacity improvement brought by RIS, and we can have the following closed-form IREE expression.
\begin{Prop}[Closed-form IREE of RIS] \label{prop:RIS}
For any given GMM based network distribution introduced by RIS, i.e., $f^{c, RIS}(\mathcal{L}, \tau)(\mathcal{L}) = \sum_{n=1}^{N} \omega^{r}_{n} \mathcal{N}(\mathcal{L}^{r}_{n}, \mathbf{\Sigma}^{r}_{n})$, the closed-form IREE of RIS configuration is given by,
\begin{eqnarray} \label{eqn:ris_iree}
\eta^{RIS}_{IREE} &=& \frac{ \min \{ C_{Tot} + \Delta C^{RIS}_{Tot}, D_{Tot}\} }{P_{Tot}}\nonumber \\
&& \times \left[1 - \xi\big( C_{Tot}, \Delta C^{RIS}_{Tot} \big) \right],
\end{eqnarray}
where $\xi\big(C_{Tot}, \Delta C^{RIS}_{Tot}\big)$ is the corresponding JS divergence given by
\begin{eqnarray}
\xi\big(C_{Tot}, \Delta C^{RIS}_{Tot}\big) 
= 1 - \frac{1}{2}\bigg[ \frac{ C_{Tot} \sum_{l = 1}^{L} \omega^c_{l} \log_2 \Big( 1 + \iota_{cd}^l \Big)  }{C_{Tot} + \Delta C^{RIS}_{Tot}}  \nonumber \\
 + \frac{ \Delta C^{RIS}_{Tot}}{C_{Tot} + \Delta C^{RIS}_{Tot}} \sum_{n = 1}^{N} \omega^r_{n} \log_2 \Big( 1 + \iota_{rd}^n \Big) + \sum_{k = 1}^{K} \omega^d_{k} \quad \nonumber \\
 \log_2 \Big ( 1 + \frac{ C_{Tot}}{C_{Tot} + \Delta C^{RIS}_{Tot}} \iota_{dc}^k + \frac{ \Delta C^{RIS}_{Tot}}{C_{Tot} + \Delta C^{RIS}_{Tot}} \iota_{dr}^k \Big) \bigg].
\end{eqnarray}
In the above equation, 
\begin{eqnarray}
\iota_{cd}^l &=& \frac{\sum_{k = 1}^{K} \omega^d_{k} I\big( \mathcal{N}(\mathcal{L}^{c}_{l}, \mathbf{\Sigma}^{c}_{l}) | \mathcal{N}(\mathcal{L}^{d}_{k}, \mathbf{\Sigma}^{d}_{k})\big)}{ \sum_{l' = 1}^{L} \omega^c_{l'} I\big( \mathcal{N}(\mathcal{L}^{c}_{l}, \mathbf{\Sigma}^{c}_{l}) | \mathcal{N}(\mathcal{L}^{c}_{l'}, \mathbf{\Sigma}^{c}_{l'})\big)}, \\
\iota_{dc}^k &=& \frac{\sum_{l = 1}^{L} \omega^c_{l} I\big( \mathcal{N}(\mathcal{L}^d_{k}, \mathbf{\Sigma}^d_{k}) | \mathcal{N}(\mathcal{L}^c_{l}, \mathbf{\Sigma}^c_{l})\big) }{ \sum_{k' = 1}^{K} \omega^d_{k'} I\big( \mathcal{N}(\mathcal{L}^d_{k}, \mathbf{\Sigma}^d_{k}) | \mathcal{N}(\mathcal{L}^d_{k'}, \mathbf{\Sigma}^d_{k'})\big) }, \\
\iota_{rd}^n &=& \frac{  \sum_{k = 1}^{K} \omega^d_{k} I\big( \mathcal{N}(\mathcal{L}^{r}_{n}, \mathbf{\Sigma}^{r}_{n}) | \mathcal{N}(\mathcal{L}^{d}_{k}, \mathbf{\Sigma}^{d}_{k})\big)}{ \sum_{n' = 1}^{N} \omega^r_{n'} I\big( \mathcal{N}(\mathcal{L}^{r}_{n}, \mathbf{\Sigma}^{r}_{n}) | \mathcal{N}(\mathcal{L}^{r}_{n'}, \mathbf{\Sigma}^{r}_{n'})\big)}, \\
\iota_{dr}^k &=& \frac{\sum_{n = 1}^{N} \omega^r_{n} I\big( \mathcal{N}(\mathcal{L}^d_{k}, \mathbf{\Sigma}^d_{k}) | \mathcal{N}(\mathcal{L}^r_{n}, \mathbf{\Sigma}^r_{n})\big) }{ \sum_{k' = 1}^{K} \omega^d_{k'} I\big( \mathcal{N}(\mathcal{L}^d_{k}, \mathbf{\Sigma}^d_{k}) | \mathcal{N}(\mathcal{L}^d_{k'}, \mathbf{\Sigma}^d_{k'})\big) }.
\end{eqnarray}
\end{Prop}
\IEEEproof
Please refer to Appendix~\ref{appendix:ris} for the proof.
\endIEEEproof

From Proposition~\ref{prop:RIS}, we show that the IREE of RIS configuration can be improved, when $\partial \eta^{RIS}_{IREE}/ \partial \Delta C^{RIS}_{Tot}$ is greater than zero. By applying the chain rule, we realize that the improvement of IREE for RIS architecture is two-fold. For one thing, the IREE can be increased by the improvement of network capacity, $\Delta C^{RIS}_{Tot}$, due to a better propagation environment provided by RIS. For another, a better matching between the network capacity and the corresponding traffic demand, i.e., the reduction of $\xi\big(C_{Tot}, \Delta C^{RIS}_{Tot}\big)$, is also preferable for the IREE improvement. This part is significantly different from the conventional energy efficient scheme, where the mismatch between the network capacity and the traffic demand is ignored. Specifically, when the network capacity can perfectly match the local traffic demand through RIS, the IREE is able to reach the maximum value as given by Proposition~\ref{prop:RIS}.

\begin{table*} [t] 
\centering 
\caption{Simulation parameters}  
\label{tab:simu_para}
\footnotesize
\begin{tabular}{c c c c}  
\toprule
Parameters & Terrestrial BS & Airborne BS (UAV) & Satellite BS (LEO) \\
\midrule
User density &  \multicolumn{3}{c}{ $10^6$ per cubic  kilometer } \\
\midrule
Frequency &  \multicolumn{3}{c}{$2$ GHz (L/S band)} \\
\midrule
System bandwidth & \multicolumn{3}{c}{$20$ MHz}  \\
\midrule 
Noise power spectral density &
\multicolumn{3}{c}{$-174$ dBm/Hz }  \\
\midrule  
Transmit power & $35$ dBm & $35$ dBm & $35$ dBm   \\
\midrule  
Circuit power & $1000$ mW & $4000$ mW & $1000$ mW  \\
\midrule  
\tabincell{c}{ Pathloss model \\ ($d$ in m) }  & 
\tabincell{c}{  LoS: $35+38 \log_{10}(d)$ dB \cite{6334508} \\  NLoS: $35+40 \log_{10}(d)$ dB } & 
$78 +20 \log_{10}(d)$ dB \cite{khuwaja2018survey} &
$148$ dB \cite{ruan2018energy}  \\
\midrule 
Deployment height  & 
$35$ m  & $100$ m & $300~800$ km \\
\midrule 
Antenna gain  & $0$ dBi &
$0$ dBi & $12$ dBi \\
\midrule 
CapEx  &  $1$ $\mathbf{M} \$ $/year &
$0.001$ $\$ $/hr \cite{lucic2020generalized} & $22.5$ $\mathbf{M} \$ $/year  \cite{infolysis2017comparative} \\
\midrule 
OpEx factor  & $0.1$ \$/kWh &
$0.00738$ $\$ $/hr \cite{lucic2020generalized} & / \\
\bottomrule
\end{tabular}  
\end{table*} 

\subsection{SAGIN Architecture}

Different from the RIS case, the SAGIN architecture provides seamless and reliable wireless services via power radiation from airborne and satellite BSs. Denote $\Delta C^{A/S}_{Tot}$ and $\Delta P^{A/S}_{Tot}$ to be the additional capacity improvement and power consumption brought by airborne or satellite BSs, and the IREE of SAGIN architecture can be summarized in the following proposition.

\begin{Prop}[IREE Bound of SAGIN] \label{prop:SAGIN}
The lower bound of IREE for the SAGIN architecture, $\bar{\eta}^{SAGIN}_{IREE}$, is given by,
\begin{eqnarray} \label{eqn:sagin_iree}
\bar{\eta}^{SAGIN}_{IREE} & = & \frac{ \min \left \{ 1, \frac{D_{Tot}}{C_{Tot} + \Delta C^{A/S}_{Tot}} \right \}}{P^T_{Tot} + \Delta P^{A,S}_{Tot} (\Delta C^{A/S}_{Tot})} \times \bigg[ C^T_{Tot} \left(1 - \xi^{T}\right) \nonumber \\
&& + \Delta C^{A/S}_{Tot} \left(1 - \xi^{A/S}(\Delta C^{A/S}_{Tot}) \right) \bigg],
\end{eqnarray}
where $\xi^{A/S}(\Delta C^{A/S}_{Tot})$ is the JS divergence improvement brought by airborne or satellite BSs and $\xi^{T}$ is the JS divergence of terrestrial BSs.
\end{Prop}
\IEEEproof
Please refer to Appendix~\ref{appendix:sagin} for the proof.
\endIEEEproof

From Proposition~\ref{prop:SAGIN}, we can figure out that the IREE improvement relies on both the capacity enhancement and the JS divergence refinement, which is also different from the conventional energy efficient strategies. Specifically, the potential IREE benefits of the SAGIN architecture can be summarized below.
\begin{itemize}
    \item {\em Capacity Enhancement}. For underserved areas with sparse traffic demands or emergency communications scenarios, extensively deploying terrestrial networks will result in significant power consumption with marginal IREE improvement. In these cases, airborne and satellite BSs are more preferable to achieve better IREE performance. This is because unlike conventional terrestrial BSs, the required transmission power budget for airborne and satellite BSs do not need to scale with the coverage areas, while some limited power consumption with the ratio equal to $\Delta P^{A/S}_{Tot}/P^T_{Tot}$, can bring a great portion of capacity enhancement as given by $\Delta C^{A/S}_{Tot}/C^T_{Tot}$.
    \item {\em Divergence Refinement}. Another important factor to improve the IREE performance of SAGIN architecture is to increase the ratio between $(1 - \xi^{A/S}(\Delta C^{A/S}_{Tot}))$ and $(1 - \xi^{T})$. For rural areas or emergency communication scenarios, the mismatch between the terrestrial network capacity and the traffic demand often happens, which results in sufficiently small $(1 - \xi^{T})$. In these cases, we can use airborne and satellite BSs to fit the underserved traffic demands and get a large value of $(1 - \xi^{A/S}(\Delta C^{A/S}_{Tot}))$.  
\end{itemize}

In the practical network deployment, the above two effects may conflict with each other. For example, conventional energy efficient schemes \cite{ruan2018energy} to enhance the network capacity will lead to the JS divergence loss and the IREE degradation, when the available traffic demand is low. Therefore, to guarantee an improved IREE performance, we need to balance the above two effects and make sure that the partial gradient of $\partial \bar{\eta}^{SAGIN}_{IREE}/ \partial \Delta C^{RIS}_{Tot}$ is positive.

\section{Numerical results} \label{sect:num_res}

In this section, we provide some numerical results to demonstrate the IREE based green trade-offs as well as the impacts on energy efficient 6G network design. For illustration purpose, we choose the UAV BS to be the airborne communication platform and the satellite BS is operating on low earth orbits (LEO) with the altitudes ranging from 300 to 800 kilometers. In the following simulations, we choose to evaluate a cube area with edge length equal to 1 kilometer, and two different types of wireless traffics are considered. The first type is generated from a standard 3D Gaussian distribution with mean $\mathcal{L}^{d} = (300, 700, 10)$ and covariance matrix $\mathbf{\Sigma}^d = \sigma^d \cdot \mathbf{I}_3$, where $\mathbf{I}_3$ is the identity matrix of order 3. The second type is generated from 2-order GMM, where $\omega^d_1 = 1/11$, $\omega^d_2 = 10/11$,  $\mathcal{L}^{d}_{1} = (300, 700, 200)$, $\mathcal{L}^{d}_{2} = (500, 500, 500)$ and $\Sigma^d_1 = 1.6 \times 10^{5} \cdot \mathbf{I}_3$, $\Sigma^d_2 = 1 \times 10^{6} \cdot \mathbf{I}_3$. The adopted simulation parameters, unless otherwise specified, are listed in Table~\ref{tab:simu_para}.

\subsection{IREE based Green Trade-offs}

\begin{figure*}[t]
\centering
\subfigure[]{
\begin{minipage}[c]{0.45\linewidth}
\centering
\includegraphics[height=7.5cm,width=7.5cm]{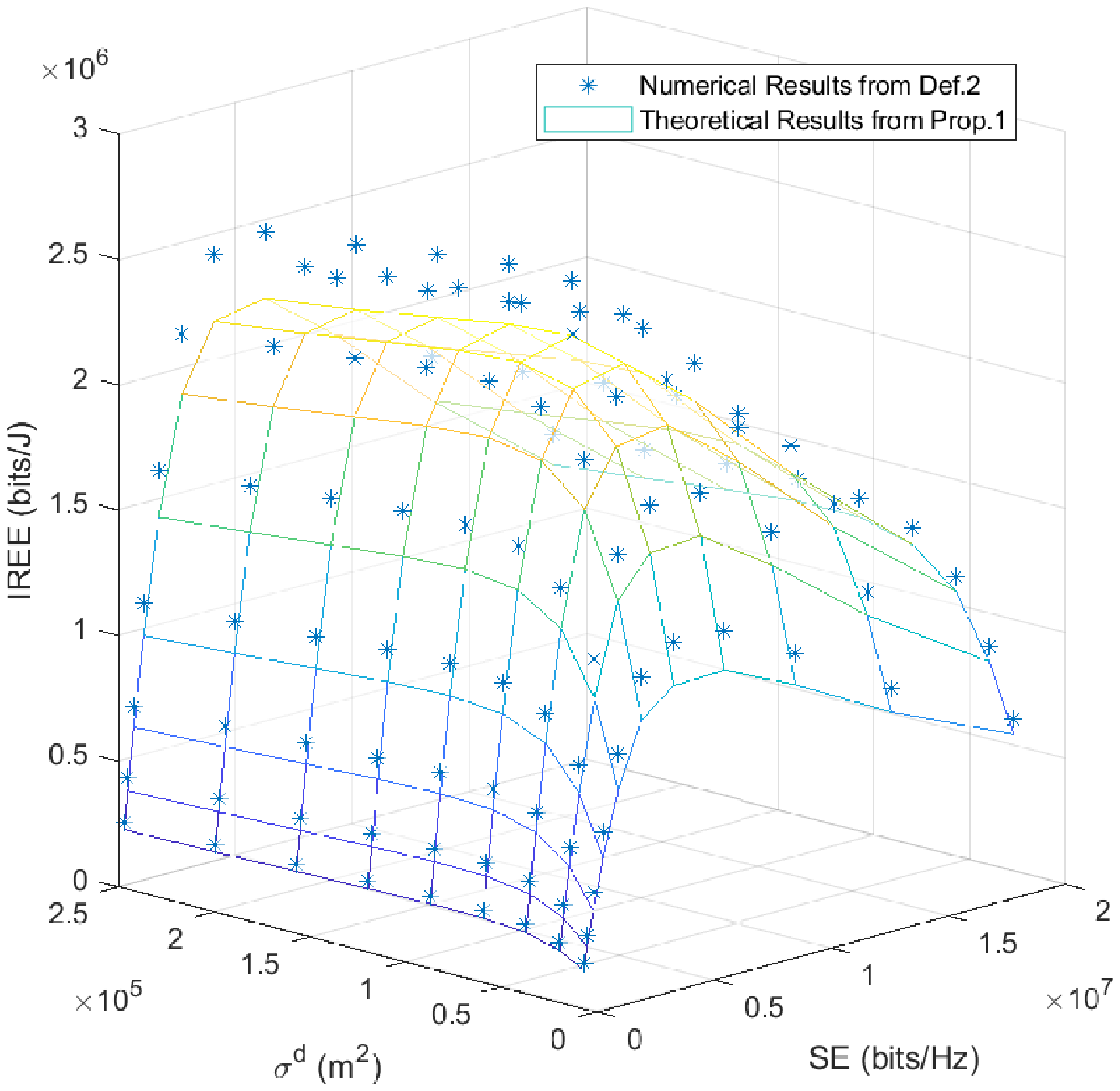}
\label{fig:iree_se}
% \vspace*{3pt}
\end{minipage}}
\subfigure[]{
\begin{minipage}[c]{0.45\linewidth}
\centering
\includegraphics[height=7.5cm,width=7.5cm]{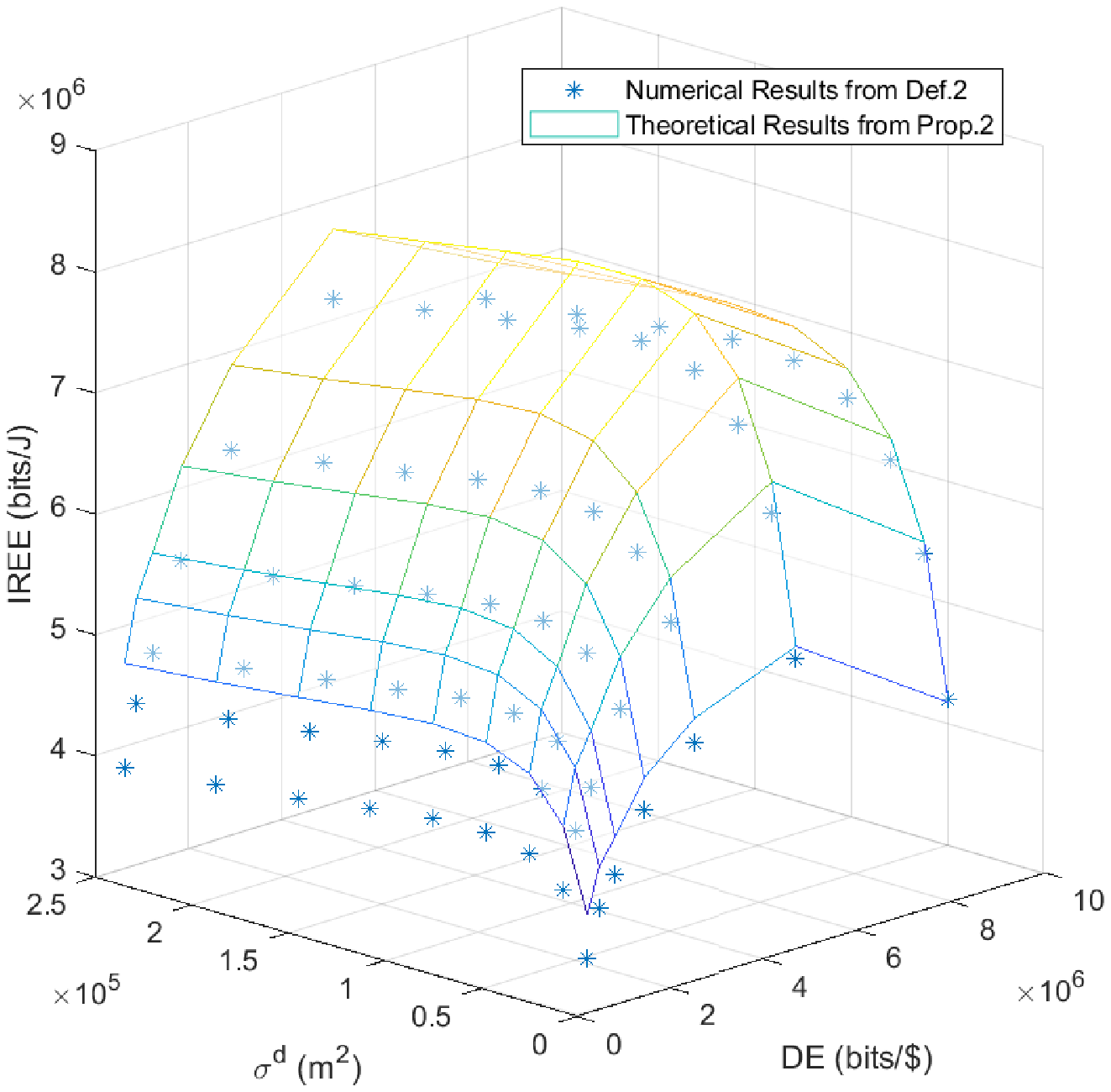}
\label{fig:iree_de}
% \vspace*{3pt}
\end{minipage}}
% \vspace*{1pt}
\caption{Analytical and numerical results for the SE-IREE and DE-IREE trade-off relations. Although for any given $\sigma^{d}$, the SE-IREE and DE-IREE trade-off relations still follow the ``bell'' shape, a dynamic variation across different traffic profiles, i.e., different values of $\sigma^{d}$, can be observed.}
\label{fig:iree_se_de}
\end{figure*}

To verify the derived SE-IREE and DE-IREE trade-off relations, we plot the achieved IREE based on Definition~\ref{def:IREE} in Fig.~\ref{fig:iree_se_de}, where different traffic profiles are generated by varying the value of $\sigma^{d}$.

In Fig.~\ref{fig:iree_se}, we generate different SE values by varying the transmit power of the terrestrial BS and the covariance matrix $\mathbf{\Sigma}^d$. As shown in Fig.~\ref{fig:iree_se}, the achieved IREE value can be well predicted by the closed-form SE-IREE trade-off relation from Proposition~\ref{prop:se_iree}. Although the SE-IREE trade-off relation still follows the ``bell'' shape, we can observe a dynamic variation across different traffic variations, and the optimal IREE is achieved when the provided network capacity perfectly matches the wireless traffic distribution.

In Fig.~\ref{fig:iree_de}, we increase the number of terrestrial BSs to generate different DE values. A trade-off relation between DE and IREE can be observed from Fig.~\ref{fig:iree_de}, which is consistent with the conventional DE-EE relation as derived in \cite{5503900}. However, the achieved IREE value follows the DE-IREE trade-off relation as described by Proposition~\ref{prop:de_iree}, which is shown to vary with the traffic distribution as well. This is because the IREE metric takes the wireless traffic demand into consideration, and the IREE degradation happens when there is insufficient traffic demand for some locations.

\subsection{Energy Efficient 6G Network Design}

In order to provide guidelines for energy efficient 6G network design, we numerically plot the JS divergences and the corresponding IREE values under two types of wireless traffics, where $\sigma^{d}$ is equal to $1\times 10^{4}$. RIS, airborne, and satellite BSs are deployed or moving in the horizontal direction as shown in Fig.~\ref{fig:scenario}.

\begin{figure}[t]
\centering
\includegraphics[width=3.3in]{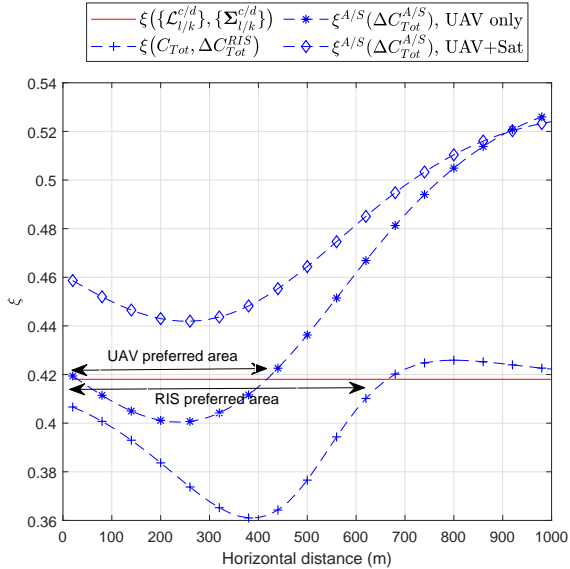}
\caption{The values of JS divergence versus the location of RIS, SAGIN with UAV BS only and UAV plus satellite BSs under a standard 3D Gaussian distribution with mean $\mathcal{L}^d = (300, 700, 10)$ and covariance matrix $\Sigma^{d} = 1 \times 10^4 \cdot \mathbf{I}_3$.}
\label{fig:NetDesign_JS_tes}
\end{figure}

\begin{figure}[t]
\centering  
\includegraphics[width=3.4in]{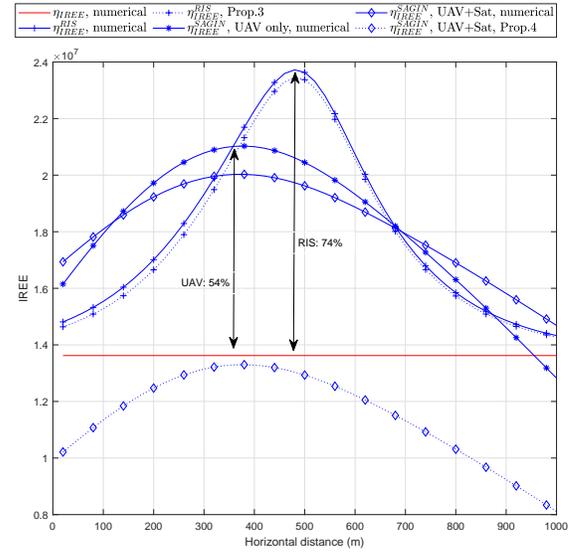}
\caption{The proposed IREE performance versus the location of RIS, SAGIN with UAV BS only and UAV plus satellite BSs under a standard 3D Gaussian distribution with mean $\mathcal{L}^d = (300, 700, 10)$ and covariance matrix $\Sigma^{d} = 1 \times 10^4 \cdot \mathbf{I}_3$.}
\label{fig:NetDesign_IREE_tes}
\end{figure}

The simulated results of the JS divergence and the IREE performance under the first type of traffic are compared in Fig.~\ref{fig:NetDesign_JS_tes} and Fig.~\ref{fig:NetDesign_IREE_tes}, respectively. As shown in Fig.~\ref{fig:NetDesign_JS_tes}, both RIS and SAGIN with UAV BSs are able to provide the JS divergence reduction for some preferred areas. Since the mean location of wireless traffic is close to the ground, RIS deployment provides more benefits in terms of the JS divergence reduction than SAGIN with UAV BSs. The similar phenomenon can also be observed for the IREE performance. As shown in Fig.~\ref{fig:NetDesign_IREE_tes}, RIS deployment achieves superior IREE improvement than SAGIN with UAV BSs, which corresponds to 74\% IREE improvement if compared with the conventional terrestrial BSs. This is partially because of the aforementioned JS divergence reduction as well as the much less energy consumption of RIS than SAGIN with UAV BSs. Moreover, the analytical results generated from Proposition~\ref{prop:RIS} and Proposition~\ref{prop:SAGIN} provide a useful reference for the numerical evaluations of the proposed IREE metric, which is of great help for us to design optimal deployment strategies.

\begin{figure}[t]
\centering
\includegraphics[width=3.3in]{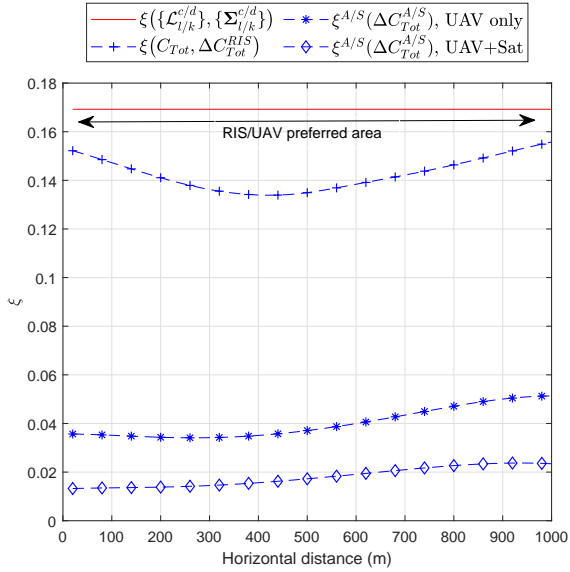}
\caption{The values of JS divergence versus the location of RIS, SAGIN with UAV BS only and UAV plus satellite BSs under a 2-order GMM, where $\omega^d_1 = 1/11$, $\omega^d_2 = 10/11$,  $\mathcal{L}^{d}_{1} = (300, 700, 200)$, $\mathcal{L}^{d}_{2} = (500, 500, 500)$ and $\Sigma^d_1 = 1.6 \times 10^{5} \cdot \mathbf{I}_3$, $\Sigma^d_2 = 1 \times 10^{6} \cdot \mathbf{I}_3$. }
\label{fig:NetDesign_JS_air}
\end{figure}

\begin{figure}[t]
\centering  
\includegraphics[width=3.4in]{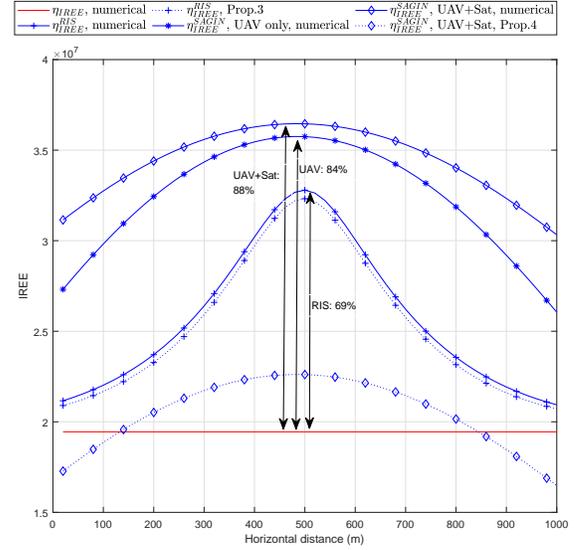}
\caption{The proposed IREE performance versus the location of RIS, SAGIN with UAV BS only and UAV plus satellite BSs under a 2-order GMM, where $\omega^d_1 = 1/11$, $\omega^d_2 = 10/11$,  $\mathcal{L}^{d}_{1} = (300, 700, 200)$, $\mathcal{L}^{d}_{2} = (500, 500, 500)$ and $\Sigma^d_1 = 1.6 \times 10^{5} \cdot \mathbf{I}_3$, $\Sigma^d_2 = 1 \times 10^{6} \cdot \mathbf{I}_3$.}
\label{fig:NetDesign_IREE_air}
\end{figure}

For the second type of traffic, the JS divergence and the IREE performance are compared in Fig.~\ref{fig:NetDesign_JS_air} and Fig.~\ref{fig:NetDesign_IREE_air}, respectively. As shown in Fig.~\ref{fig:NetDesign_JS_air}, all types of BSs are able to provide the JS divergence reduction. Since the mean location of wireless traffic is far from the ground, the UAV and satellite BSs are more preferable in this case, which is quite different from the previous settings. In Fig.~\ref{fig:NetDesign_IREE_air}, we show that the maximum IREE performance improvements for RIS and SAGIN networks with UAV only as well as UAV plus satellite BSs can reach to 69\%, 84\%, and 88\%, respectively, which also aligns with the analytical results from Proposition~\ref{prop:RIS} and Proposition~\ref{prop:SAGIN}.

\section{Conclusion} \label{sect:conc}

In this paper, we propose a novel IREE metric to deal with the dynamic network capacity and traffic variations. By integrating the 3D data traffic and capacity distribution, the proposed IREE is able to describe the mismatched traffic demand and network deployment from the EE perspective. On top of that, the IREE based fundamental green trade-offs are investigated, which is shown to be significantly different from the conventional EE based trade-offs. In addition, we apply the proposed IREE metric to investigate the potential 6G network architectures, including both RIS and SAGIN. From the analytical as well as numerical results, we show that the energy efficient strategies for 6G networks need to adaptively and dynamically provide the network capacity for some underserved areas and emergency scenarios rather than directly deploying terrestrial BSs.

\newpage

\begin{appendices} 

\section{Derivation from IEE to IREE} \label{appendix:def}

From the definition of IEE \eqref{def:iee}, we have,
\begin{eqnarray} 
&&\int_{\tau}\iiint_{\mathcal{A}}\min\{C_{T}(\mathcal{L}, \tau), D_{T}(\mathcal{L}, \tau)\} \textrm{d}\mathcal{L} \textrm{d}\tau \nonumber \\
&=& \int_{\tau}\iiint_{\mathcal{A}} C_T(\mathcal{L}, \tau) \mathcal{I} \left  [ f^{c}_N(\mathcal{L}, \tau) \leq  f^{d}_N(\mathcal{L}, \tau) \right ] \nonumber \\
&& + D_T(\mathcal{L}, \tau) \mathcal{I} \left  [f^{c}_N(\mathcal{L}, \tau) > f^{d}_N(\mathcal{L}, \tau) \right] \textrm{d}\mathcal{L} \textrm{d}\tau \nonumber \\
&=& \int_{\tau}\iiint_{\mathcal{A}} \frac{1}{2} C_T(\mathcal{L}, \tau) \Big [ \mathcal{I} \left  [ f^{c}_N(\mathcal{L}, \tau) \leq  f^{d}_N(\mathcal{L}, \tau) \right ] \nonumber \\
&&+ \frac{f^{d}_N (\mathcal{L}, \tau) }{f^{c}_N (\mathcal{L}, \tau) } \mathcal{I} \left  [f^{c}_N(\mathcal{L}, \tau) > f^{d}_N(\mathcal{L}, \tau) \right] \Big] \nonumber \\
&& + \frac{1}{2} D_T(\mathcal{L}, \tau) \Big [ \mathcal{I} \left  [ f^{d}_N(\mathcal{L}, \tau) \leq  f^{c}_N(\mathcal{L}, \tau) \right ]  \nonumber \\
&&+ \frac{f^{c}_N (\mathcal{L}, \tau) }{f^{d}_N (\mathcal{L}, \tau) } \mathcal{I} \left  [f^{d}_N(\mathcal{L}, \tau) > f^{c}_N(\mathcal{L}, \tau) \right] \Big]
\textrm{d}\mathcal{L} \textrm{d}\tau  \\
&\overset{(a)}{=}& \max \{C_{Tot}, D_{Tot}\} \times \nonumber \\
&& \int_{\tau} \iiint_{\mathcal{A}} \frac{1}{2} f^{c}_N(\mathcal{L}, \tau) \log_2 \left[ 1 + \frac{f^{d}_N(\mathcal{L}, \tau) }{ f^{c}_N(\mathcal{L}, \tau) } \right] \nonumber \\
&&+ \frac{1}{2} f^{d}_N(\mathcal{L}, \tau) \log_2 \left[ 1 + \frac{f^{c}_N(\mathcal{L}, \tau) }{ f^{d}_N(\mathcal{L}, \tau) } \right] \textrm{d}{\mathcal{L}} \textrm{d}\tau \\
&=& \max \{C_{Tot}, D_{Tot}\} \times \nonumber \\
&& \int_{\tau} \iiint_{\mathcal{A}} \frac{1}{2} f^{c}_N(\mathcal{L}, \tau) \left [ 1- \log_2 \left[ \frac{f^{c}_N(\mathcal{L}, \tau) }{\frac{f^{c}_N(\mathcal{L}, \tau) + f^{d}_N(\mathcal{L}, \tau)}{2} } \right] \right] \nonumber \\
&& + \frac{1}{2} f^{d}_N(\mathcal{L}, \tau) \left [ 1- \log_2 \left[ \frac{f^{d}_N(\mathcal{L}, \tau) }{\frac{f^{c}_N(\mathcal{L}, \tau) + f^{d}_N(\mathcal{L}, \tau)}{2} } \right] \right] \textrm{d}{\mathcal{L}} \textrm{d}\tau \nonumber \\
&=& \max \{C_{Tot}, D_{Tot}\}\Big[ \frac{1}{2} + \frac{\min \{C_{Tot}, D_{Tot}\}}{2\max \{C_{Tot}, D_{Tot}\}} \nonumber \\
&& - \xi(f^{c}_N, f^{d}_N)  \Big]\\
&\overset{(b)}{=}& \max \{C_{Tot}, D_{Tot}\}\Big[ \frac{1}{2} + \frac{\min \{C_{Tot}, D_{Tot}\}}{2\max \{C_{Tot}, D_{Tot}\}} \nonumber \\
&& - \xi \Big ( \frac{\min \{C_{Tot}, D_{Tot}\}}{\max \{C_{Tot}, D_{Tot}\}} f^{c}+ \big (1- \frac{ \min \{C_{Tot}, D_{Tot}\}}{\max \{C_{Tot}, D_{Tot}\}} \big )  \nonumber \\
&&\times f^{c}, \frac{\min \{C_{Tot}, D_{Tot}\}}{\max \{C_{Tot}, D_{Tot}\}} f^{d} \Big)  \Big] \\
&\overset{(c)}{\geq}& \max \{C_{Tot}, D_{Tot}\}\Big[ \frac{1}{2} + \frac{\min \{C_{Tot}, D_{Tot}\}}{2\max \{C_{Tot}, D_{Tot}\}} \nonumber \\
&& - \Big[ \frac{\min \{C_{Tot}, D_{Tot}\}}{\max \{C_{Tot}, D_{Tot}\}} \xi(f^{c}, f^{d}) \nonumber \\
&& + \big (1- \frac{ \min \{C_{Tot}, D_{Tot}\}}{\max \{C_{Tot}, D_{Tot}\}} \big ) \xi(f^{c}, 0) \Big] \Big] \\
&=& \max \{C_{Tot}, D_{Tot}\}\Big[ \frac{\min \{C_{Tot}, D_{Tot}\}}{\max \{C_{Tot}, D_{Tot}\}} \nonumber \\
&& - \frac{\min \{C_{Tot}, D_{Tot}\}}{\max \{C_{Tot}, D_{Tot}\}} \xi(f^{c}, f^{d})  \Big] \nonumber \\
&=& \min \{C_{Tot}, D_{Tot}\}\Big[ 1  - \xi(f^{c}, f^{d})  \Big],
\end{eqnarray}
where $\mathcal{I} \left[ \cdot \right]$ equals to 1 if the inner condition holds, and equals to 0 otherwise. $\xi(f^{c}, f^{d})$ is the JS divergence given by,
\begin{eqnarray}
&&\xi(f^{c}, f^{d}) \nonumber \\
&=& \frac{1}{2} \int_{\tau}\iiint_{\mathcal{A}}  f^{c}(\mathcal{L}, \tau) \log_2 \left[ \frac{2 f^{c}(\mathcal{L}, \tau) }{f^{c}(\mathcal{L}, \tau) + f^{d}(\mathcal{L}, \tau) } \right] \nonumber \\
&&+ f^{d}(\mathcal{L}, \tau) \log_2 \left[ \frac{2 f^{d}(\mathcal{L},\tau)}{f^{c}(\mathcal{L}, \tau) + f^{d}(\mathcal{L}, \tau)} \right] \textrm{d}\mathcal{L} \textrm{d}\tau. \qquad 
\end{eqnarray}
$C_{Tot}$ and $D_{Tot}$ are the total amount of network capacity and wireless traffic as defined in Definition~\ref{def:IREE}. $f^{d}_N(\mathcal{L}, \tau) = \frac{D_{Tot}}{ \max \{C_{Tot}, D_{Tot}\} } f^{d}(\mathcal{L}, \tau)  $ and $f^{c}_N(\mathcal{L}, \tau) = \frac{C_{Tot}}{ \max \{C_{Tot}, D_{Tot}\} } f^{c}(\mathcal{L}, \tau)$  are the normalized network capacity and traffic distribution. 

In step (a), we use a continuous function $\log_2 \left[ 1 + \frac{f^{c}_N(\mathcal{L}, \tau) }{ f^{d}_N(\mathcal{L}, \tau) } \right]$ to measure the difference between $f^{c}_N(\mathcal{L}, \tau)$ and $f^{d}_N(\mathcal{L}, \tau)$ \footnote{In the field of information theory, the logarithmic function is usually used to represent the difference between distributions\cite{kullback1951information}.}. In step (b), we have $f^{c}_N(\mathcal{L}, \tau) = f^{c}(\mathcal{L}, \tau)$,  $f^{d}_N(\mathcal{L}, \tau) = \frac{\min \{C_{Tot}, D_{Tot}\}}{\max \{C_{Tot}, D_{Tot}\}} f^{d}(\mathcal{L}, \tau)$ for $D_{Tot} \leq C_{Tot}$, and $f^{c}_N(\mathcal{L}, \tau) = \frac{\min \{C_{Tot}, D_{Tot}\}}{\max \{C_{Tot}, D_{Tot}\}} f^{c}(\mathcal{L}, \tau)$,  $f^{d}_N(\mathcal{L}, \tau) = f^{d}(\mathcal{L}, \tau)$ for $D_{Tot} > C_{Tot}$, respectively. In step (c), we directly apply the Jensen's inequality in the derivation. With the above explanation, IREE is defined as,
\begin{eqnarray} 
\eta_{IREE} = \frac{ \min \{C_{Tot}, D_{Tot}\}\Big[ 1  - \xi(f^{c}, f^{d})  \Big] } {\int_{\tau}P_{T}(\tau)\textrm{d}\tau }.
\end{eqnarray}

\section{Proof of Proposition~\ref{prop:se_iree}} \label{appendix:se_iree}

Mathematically, the JS divergence $\xi\left(f^c, f^d \right)$ is equal to the arithmetic mean of two Kullback-Leibler (KL) divergence $\xi_{KL} \left(f^c \big | \frac{f^c + f^d }{2} \right)$ and $\xi_{KL} \left(f^d \big | \frac{f^c + f^d }{2} \right)$, i.e.

\begin{equation}
\label{eq:js_exp}
\xi\left(f^c, f^d \right) = \frac{1}{2} \left [ \xi_{KL} \left ( f^c \Big | \frac{f^c + f^d }{2} \right ) + \xi_{KL} \left ( f^d \Big | \frac{f^c + f^d }{2} \right) \right ],
\end{equation}
where $\xi_{KL}\left ( g | h  \right ) = \int_{\tau}\iiint_{\mathcal{A}}  g(\mathcal{L}, \tau) \ln \left[ \frac{ g(\mathcal{L}, \tau) }{ h(\mathcal{L}, \tau) } \right] \textrm{d}\mathcal{L} \textrm{d}\tau$. From the variational approximation method proposed in \cite{hershey2007approximating}, the closed-form approximation of $\xi_{KL} \left ( f^c \Big | \frac{f^c + f^d }{2} \right )$ and $\xi_{KL} \left ( f^d \Big | \frac{f^c + f^d }{2} \right )$ is given by:
\begin{eqnarray}
\label{eq:kl_1}
&&\xi_{KL} \left ( f^c \Big | \frac{f^c + f^d }{2} \right )  = 1 - \sum_{l = 1}^{L} \omega^c_{l} \log_2 \Bigg ( 1 + \nonumber \\
&& \frac{  \sum_{k = 1}^{K} \omega^d_{k} I\big( \mathcal{N}(\mathcal{L}^{c}_{l}, \mathbf{\Sigma}^{c}_{l}) | \mathcal{N}(\mathcal{L}^{d}_{k}, \mathbf{\Sigma}^{d}_{k})\big)}{ \sum_{l' = 1}^{L} \omega^c_{l'} I\big( \mathcal{N}(\mathcal{L}^{c}_{l}, \mathbf{\Sigma}^{c}_{l}) | \mathcal{N}(\mathcal{L}^{c}_{l'}, \mathbf{\Sigma}^{c}_{l'})\big)} \Bigg ),
\end{eqnarray}
\begin{eqnarray}
\label{eq:kl_2}
&&\xi_{KL} \left ( f^d \Big | \frac{f^c + f^d }{2} \right )  = 1 - \sum_{k = 1}^{K} \omega^d_{k} \log_2 \Bigg ( 1 + \nonumber \\
&& \frac{\sum_{l = 1}^{L} \omega^c_{l} I\big( \mathcal{N}(\mathcal{L}^d_{k}, \mathbf{\Sigma}^d_{k}) | \mathcal{N}(\mathcal{L}^c_{l}, \mathbf{\Sigma}^c_{l})\big) }{ \sum_{k' = 1}^{K} \omega^d_{k'} I\big( \mathcal{N}(\mathcal{L}^d_{k}, \mathbf{\Sigma}^d_{k}) | \mathcal{N}(\mathcal{L}^d_{k'}, \mathbf{\Sigma}^d_{k'})\big) } \Bigg ).
\end{eqnarray}
In the above expression, the exponential mutual divergence, $I(\mathcal{N}(\mathcal{L}^{g}, \mathbf{\Sigma}^{g})|\mathcal{N}(\mathcal{L}^{h}, \mathbf{\Sigma}^{h}))$, is given by
\begin{eqnarray}
&&I(\mathcal{N}(\mathcal{L}^{g}, \mathbf{\Sigma}^{g})|\mathcal{N}(\mathcal{L}^{h}, \mathbf{\Sigma}^{h})) \nonumber \\
&&= \text{exp} \Big \{-\frac{1}{2} \Big[ \ln\frac{|\mathbf{\Sigma}^h|}{|\mathbf{\Sigma}^g|} + \text{Tr}[ (\mathbf{\Sigma}^h)^{-1} \mathbf{\Sigma}^g ] \nonumber \\
&& \quad + (\mathcal{L}^h - \mathcal{L}^g)^T (\mathbf{\Sigma}^h)^{-1} (\mathcal{L}^h - \mathcal{L}^g)-3\Big] \Big \}.
\end{eqnarray}
Substituting \eqref{eq:kl_2}, \eqref{eq:kl_1} and \eqref{eq:js_exp} into \eqref{eq:se_iree_thm} completes the proof of Proposition~\ref{prop:se_iree}.

\section{Proof of Proposition~\ref{prop:RIS}} \label{appendix:ris}

The network distribution of RIS assisted network is given by
\begin{eqnarray}
f^{c, Tot}(\mathcal{L}, \tau) & = & \frac{C(\mathcal{L}, \tau) + \Delta C(\mathcal{L}, \tau) }{ C_{Tot} + \Delta C^{RIS}_{Tot}} \nonumber \\
& = & \frac{C_{Tot}}{C_{Tot} + \Delta C^{RIS}_{Tot}} f^{c}(\mathcal{L}, \tau)  + \nonumber \\
&&  \frac{\Delta C^{RIS}_{Tot} }{C_{Tot} + \Delta C^{RIS}_{Tot}} f^{c, RIS}(\mathcal{L}, \tau),
\end{eqnarray}
where $\Delta C^{RIS}_{Tot} = \int_{\tau}\iiint_{\mathcal{A}} \Delta C^{RIS}(\mathcal{L}, \tau) \textrm{d}\mathcal{L} \textrm{d}\tau$. 

For any GMM based network distribution introduced by RIS $f^{c, RIS}(\mathcal{L}, \tau)(\mathcal{L}) = \sum_{n=1}^{N} \omega^{r}_{n} \mathcal{N}(\mathcal{L}^{r}_{n}, \mathbf{\Sigma}^{r}_{n})$, we have the following close-form expression from the close-form expression of JS divergence in Proposition~\ref{prop:se_iree},
\begin{eqnarray}
&& \xi\big(C_{Tot}, \Delta C^{RIS}_{Tot}\big) \nonumber \\
&& = 1 - \frac{1}{2}\bigg[ \frac{ C_{Tot} \sum_{l = 1}^{L} \omega^c_{l} \log_2 \Big( 1 + \iota_{cd}^l \Big)  }{C_{Tot} + \Delta C^{RIS}_{Tot}}  \nonumber \\
&&  + \frac{ \Delta C^{RIS}_{Tot}}{C_{Tot} + \Delta C^{RIS}_{Tot}} \sum_{n = 1}^{N} \omega^r_{n} \log_2 \Big( 1 + \iota_{rd}^n \Big)  \nonumber \\
&& + \sum_{k = 1}^{K} \omega^d_{k} \log_2 \Big ( 1 + \frac{ C_{Tot}}{C_{Tot} + \Delta C^{RIS}_{Tot}} \iota_{dc}^k \nonumber \\
&& + \frac{ \Delta C^{RIS}_{Tot}}{C_{Tot} + \Delta C^{RIS}_{Tot}} \iota_{dr}^k \Big) \bigg],
\end{eqnarray}
where 
\begin{eqnarray}
\iota_{cd}^l &=& \frac{\sum_{k = 1}^{K} \omega^d_{k} I\big( \mathcal{N}(\mathcal{L}^{c}_{l}, \mathbf{\Sigma}^{c}_{l}) | \mathcal{N}(\mathcal{L}^{d}_{k}, \mathbf{\Sigma}^{d}_{k})\big)}{ \sum_{l' = 1}^{L} \omega^c_{l'} I\big( \mathcal{N}(\mathcal{L}^{c}_{l}, \mathbf{\Sigma}^{c}_{l}) | \mathcal{N}(\mathcal{L}^{c}_{l'}, \mathbf{\Sigma}^{c}_{l'})\big)}, \\
\iota_{dc}^k &=& \frac{\sum_{l = 1}^{L} \omega^c_{l} I\big( \mathcal{N}(\mathcal{L}^d_{k}, \mathbf{\Sigma}^d_{k}) | \mathcal{N}(\mathcal{L}^c_{l}, \mathbf{\Sigma}^c_{l})\big) }{ \sum_{k' = 1}^{K} \omega^d_{k'} I\big( \mathcal{N}(\mathcal{L}^d_{k}, \mathbf{\Sigma}^d_{k}) | \mathcal{N}(\mathcal{L}^d_{k'}, \mathbf{\Sigma}^d_{k'})\big) }, \\
\iota_{rd}^n &=& \frac{  \sum_{k = 1}^{K} \omega^d_{k} I\big( \mathcal{N}(\mathcal{L}^{r}_{n}, \mathbf{\Sigma}^{r}_{n}) | \mathcal{N}(\mathcal{L}^{d}_{k}, \mathbf{\Sigma}^{d}_{k})\big)}{ \sum_{n' = 1}^{N} \omega^r_{n'} I\big( \mathcal{N}(\mathcal{L}^{r}_{n}, \mathbf{\Sigma}^{r}_{n}) | \mathcal{N}(\mathcal{L}^{r}_{n'}, \mathbf{\Sigma}^{r}_{n'})\big)}, \\
\iota_{dr}^k &=& \frac{\sum_{n = 1}^{N} \omega^r_{n} I\big( \mathcal{N}(\mathcal{L}^d_{k}, \mathbf{\Sigma}^d_{k}) | \mathcal{N}(\mathcal{L}^r_{n}, \mathbf{\Sigma}^r_{n})\big) }{ \sum_{k' = 1}^{K} \omega^d_{k'} I\big( \mathcal{N}(\mathcal{L}^d_{k}, \mathbf{\Sigma}^d_{k}) | \mathcal{N}(\mathcal{L}^d_{k'}, \mathbf{\Sigma}^d_{k'})\big) }.
\end{eqnarray}
Substitute the above equation into \eqref{eq:def_iree}, we have  Proposition~\ref{prop:RIS}. 

\section{Proof of Proposition~\ref{prop:SAGIN}} \label{appendix:sagin}

The network distribution of SAGIN can be rewrite as
\begin{eqnarray}
f^{c, Tot}(\mathcal{L}, \tau) &=& \frac{ C^T_{Tot}}{C^T_{Tot} + \Delta C^{A,S}_{Tot}} f^{c, T}(\mathcal{L}, \tau) \nonumber \\
&& + \frac{ \Delta C^{A,S}_{Tot}}{C^T_{Tot} + \Delta C^{A,S}_{Tot}} f^{c, A/S}(\mathcal{L}, \tau),
\end{eqnarray}
where the corresponding distributions $f^{c, T} (\mathcal{L}, \tau) = \frac{ C^{T}(\mathcal{L}, \tau) }{C^{T}_{Tot}}$, $ C^{T}_{Tot} = \int_{\tau}\iiint_{\mathcal{A}} C^{T}(\mathcal{L}, \tau) \textrm{d}\mathcal{L} \textrm{d}\tau$, and  $f^{c, A/S} (\mathcal{L}, \tau) = \frac{ \Delta C^{A/S}(\mathcal{L}, \tau) }{\Delta C^{A/S}_{Tot}}$, $\Delta C^{A/S}_{Tot} = \int_{\tau}\iiint_{\mathcal{A}} \Delta C^{A/S}(\mathcal{L}, \tau) \textrm{d}\mathcal{L} \textrm{d}\tau$.

Since $\xi[f^{c}, f^{d}]$ is convex in the pair of probability density functions $(f^{c}, f^{d})$, the upper bound can be obtained by applying Jensen's inequality as follows
\begin{eqnarray}
\label{eq:sagin_rdf}
&&\xi[f^{c, Tot}(\mathcal{L}), f^{d}(\mathcal{L})] \nonumber \\
&&= \xi \left[ \frac{ C^T_{Tot} f^{c, T}(\mathcal{L}) +\Delta C^{A,S}_{Tot}  f^{c, A/S}(\mathcal{L}) }{C^T_{Tot} + \Delta C^{A,S}_{Tot}}, f^{d}(\mathcal{L}) \right] \nonumber \\
&& \leq \frac{ C^T_{Tot} }{C^T_{Tot} + \Delta C^{A,S}_{Tot}} \xi^{T} +  \nonumber \\
&& \frac{ \Delta C^{A,S}_{Tot} }{C^T_{Tot} + \Delta C^{A,S}_{Tot}} \xi^{A/S}(\Delta C^{A/S}_{Tot}).
\end{eqnarray} 
Substitute \eqref{eq:sagin_rdf} into \eqref{eq:def_iree}, we have Proposition~\ref{prop:SAGIN}.
\end{appendices} 

\bibliographystyle{IEEEtran}
\bibliography{IEEEfull,references}

\begin{IEEEbiography}[{\includegraphics[width=1in,height=1.25in,clip,keepaspectratio]{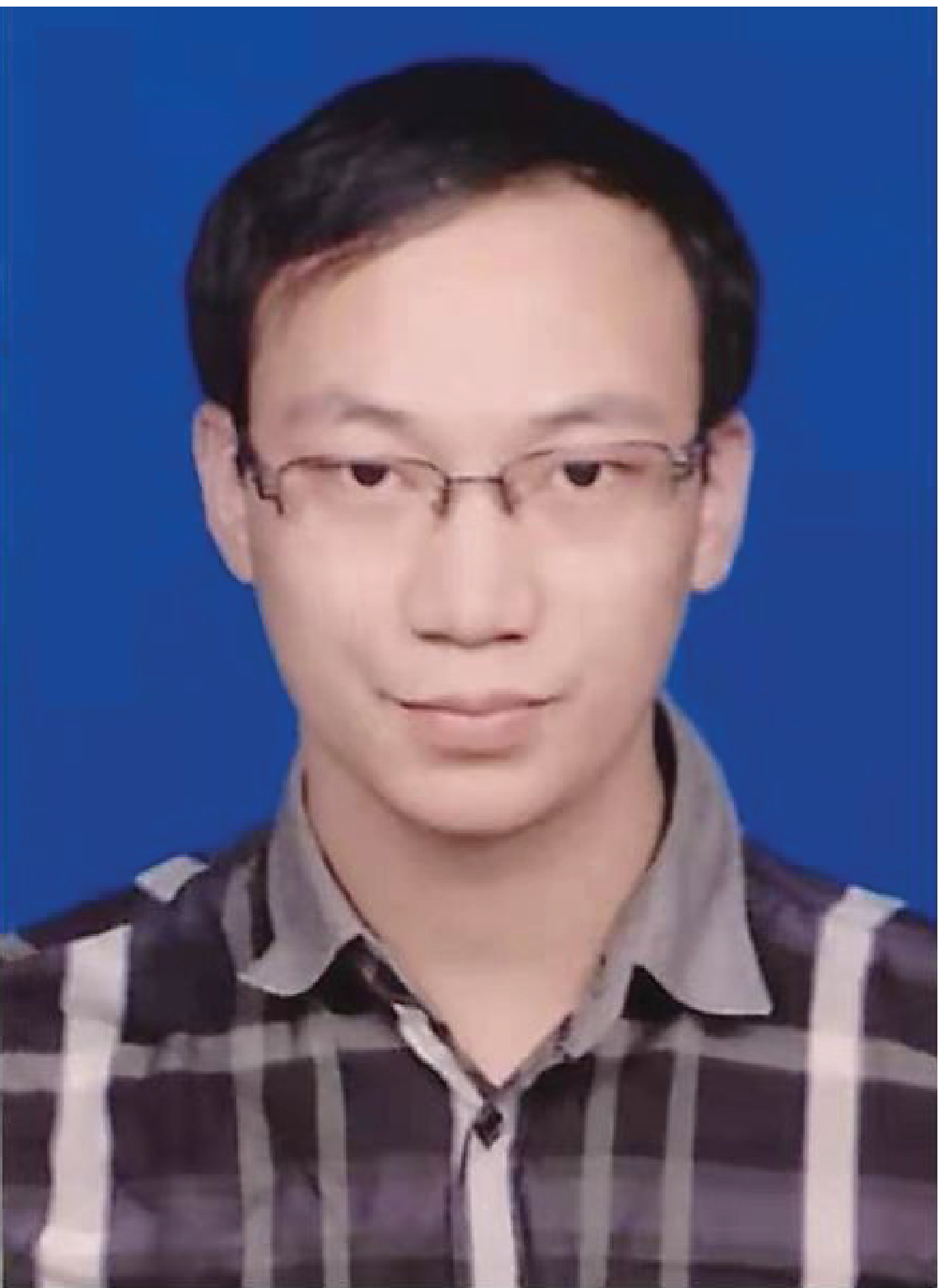}}]{Tao Yu}
received the B.E. and M.E. degrees from the School of Communication and Information Engineering, Shanghai University, in 2018 and 2021, respectively, where he is currently pursuing the Ph.D. degree with the School of Communication and Information Engineering. His research fields include energy-efficient communication networks, machine learning, and deep learning in the PHY layer.
\end{IEEEbiography}

\begin{IEEEbiography}[{\includegraphics[width=1in,height=1.25in,clip,keepaspectratio]{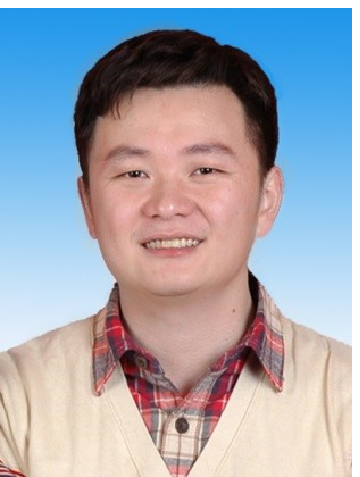}}]{Shunqing Zhang}
(Senior Member, IEEE) received the B.S. degree from the Department of Microelectronics, Fudan University, Shanghai, China, in 2005, and the Ph.D. degree from the Department of Electrical and Computer Engineering, Hong Kong University of Science and Technology, Hong Kong, in 2009.

He was with the Communication Technologies Laboratory, Huawei Technologies, as a Research Engineer and then a Senior Research Engineer from 2009 to 2014, and a Senior Research Scientist of Intel Collaborative Research Institute on Mobile Networking and Computing, Intel Labs from 2015 to 2017. Since 2017, he has been with the School of Communication and Information Engineering, Shanghai University, Shanghai, China, as a Full Professor. His current research interests include energy efficient 5G/5G+ communication networks, hybrid computing platform, and joint radio frequency and baseband design. He has published over 60 peer-reviewed journal and conference papers, as well as over 50 granted patents. He has received the National Young 1000-Talents Program and won the paper award for Advances in Communications from IEEE Communications Society in 2017.
\end{IEEEbiography}

\begin{IEEEbiography}[{\includegraphics[width=1in,height=1.25in,clip,keepaspectratio]{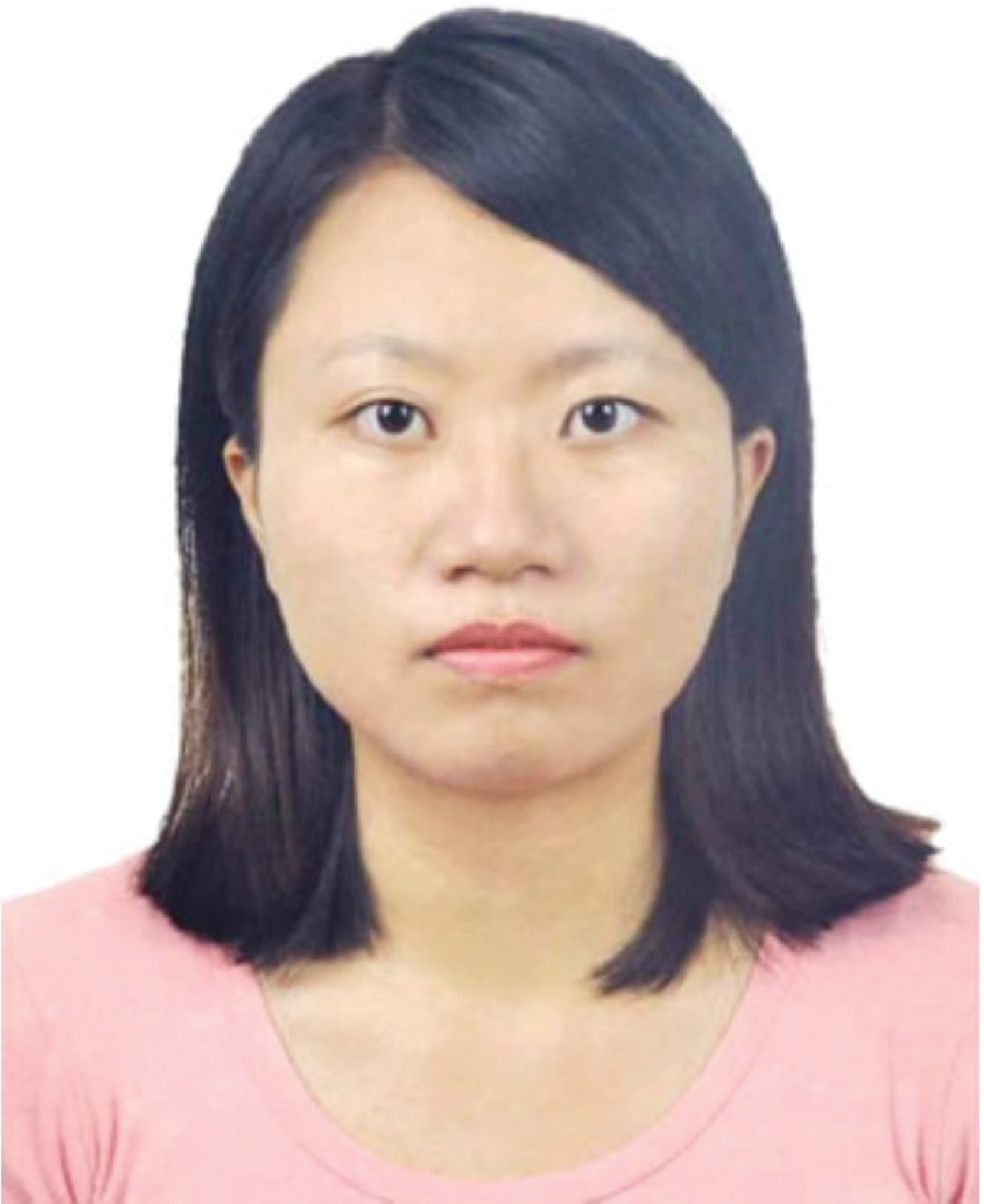}}]{Xiaojing Chen}
(Member, IEEE) received the B.E. degree in communication science and engineering and the Ph.D. degree in electromagnetic field and microwave technology from Fudan University, China, in 2013 and 2018, respectively, and the Ph.D. degree in engineering from Macquarie University, Australia, in 2019. She is currently a Lecturer with Shanghai University, China. Her research interests include wireless communications, energy-efficient communications, stochastic network optimization, and network functions virtualization.
\end{IEEEbiography}

\begin{IEEEbiography}[{\includegraphics[width=1in,height=1.25in,clip,keepaspectratio]{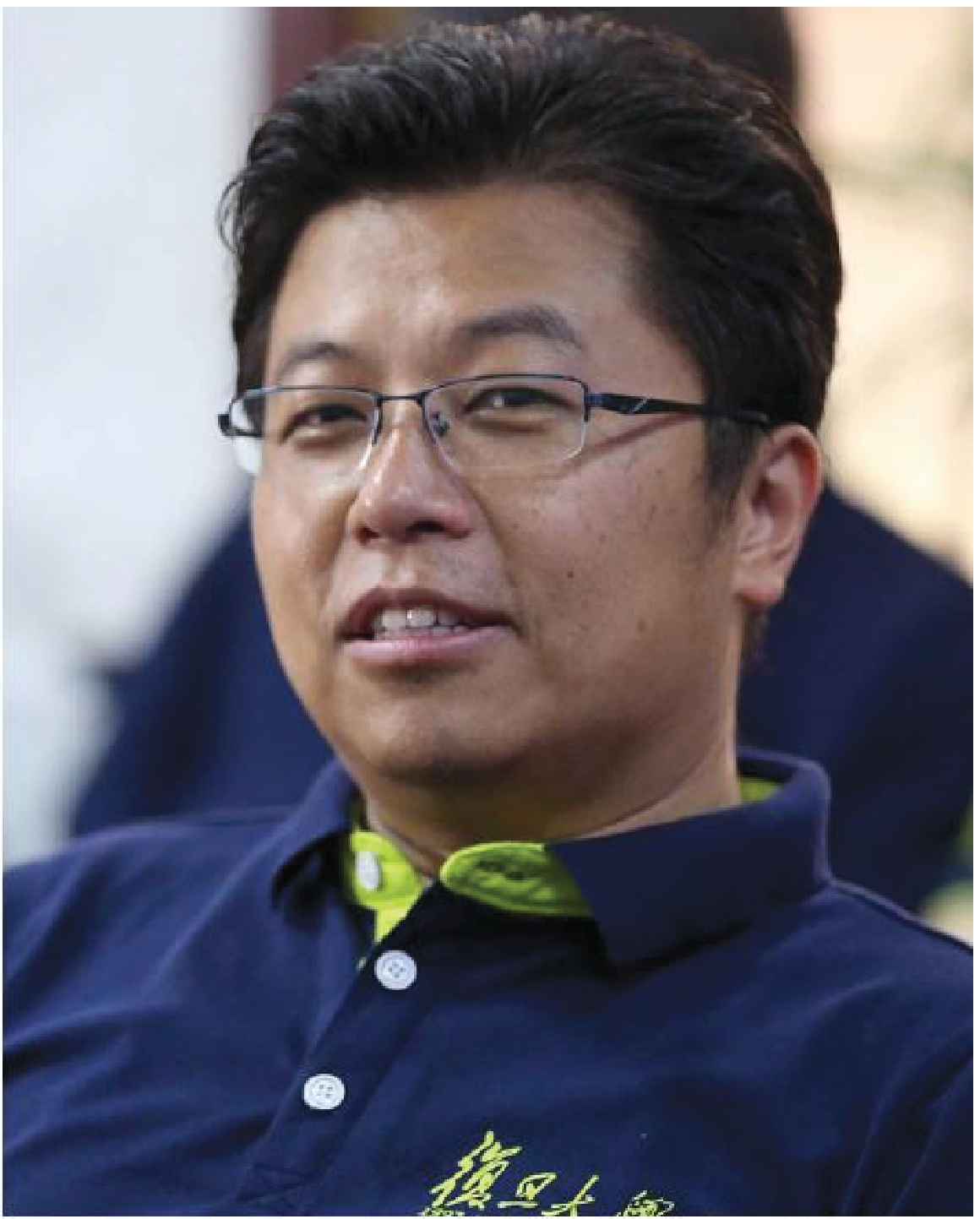}}]{Xin Wang}
(Senior Member, IEEE) received the B.Sc. and M.Sc. degrees in electrical engineering from Fudan University, Shanghai, China, in 1997 and 2000, respectively, and the Ph.D. degree in electrical engineering from Auburn University, Auburn, AL, USA, in 2004.

From September 2004 to August 2006, he was a Post-Doctoral Research Associate with the Department of Electrical and Computer Engineering, University of Minnesota, Minneapolis. In August 2006, he joined the Department of Electrical Engineering, Florida Atlantic University, Boca Raton, FL, USA, as an Assistant Professor, then was promoted to a Tenured Associate Professor in 2010. He is currently a Distinguished Professor and the Chair of the Department of Communication Science and Engineering, Fudan University. His research interests include stochastic network optimization, energy-efficient communications, cross-layer design, and signal processing for communications. He is an IEEE Distinguished Lecturer of the Vehicular Technology Society. He served as an Associate Editor for IEEE Transactions on Signal Processing and IEEE Signal Processing Letters. He also served as an Editor for IEEE Transactions on Vehicular Technology. He currently serves as a Senior Area Editor for IEEE Transactions on Signal Processing and an Editor for IEEE Transactions on Wireless Communications.
\end{IEEEbiography}

\end{document}